\documentstyle[preprint,aps]{revtex}
\begin{document}
\draft
\preprint{\hbox to \hsize{\hfil\vtop{\hbox{IASSNS-HEP-98/46}
\hbox{August, 1998}}}}

\title{Model for Particle Masses, Flavor Mixing, and CP Violation,\\
Based on Spontaneously Broken Discrete Chiral Symmetry\\
as the Origin of Families\\
}
\author{Stephen L. Adler\\}
\address{
Institute for Advanced Study\\
Princeton, NJ 08540\\
}
\maketitle

\leftline{\it To be published in Phys. Rev. D}
\leftline{\it Send correspondence to:}

\leftline{Stephen L. Adler}
\leftline{Institute for Advanced Study}
\leftline{Olden Lane, Princeton, NJ 08540}
\leftline{Phone 609-734-8051;  609-924-8399; email adler@ias.edu}

\begin{abstract}
We construct extensions of the standard model based on the hypothesis 
that the Higgs bosons also exhibit a family structure, and 
that the flavor weak eigenstates in the three families are 
distinguished by a discrete $Z_6$ chiral symmetry that is spontaneously 
broken by the Higgs sector.  We study in detail at the tree level 
models with three 
Higgs doublets, and with six Higgs doublets comprising 
two weakly coupled sets of three.  
In a leading approximation of $S_3$ cyclic permutation symmetry the 
three Higgs model gives a ``democratic'' mass matrix of rank one, while 
the six Higgs model gives either a rank one mass matrix, or in the case 
when it spontaneously violates CP, a rank two mass matrix corresponding to 
nonzero second family masses.  In both models, the CKM matrix is exactly 
unity in leading approximation.  
Allowing small explicit violations of cyclic permutation symmetry 
generates small first family masses in the six Higgs model, and first and 
second family masses in the three Higgs model, and 
gives a non-trivial CKM matrix in 
which the mixings of the first and second family quarks are naturally 
larger than mixings involving the third family.  Complete numerical fits 
are given for both models, flavor changing neutral current constraints are 
discussed in detail, and the issues of unification of couplings and  
neutrino masses are addressed.  On a 
technical level, our analysis uses the theory of circulant 
and retrocirculant matrices, the relevant parts of which are reviewed.    
\end{abstract}

\section*{I.~~Introduction}

It has long been recognized that the hierarchical structures of the 
family mass spectra, with their large third family masses, and of the CKM 
mixing matrix, with its suppressed third family mixings, may have a common
dynamical origin.  In particular, several authors [1] have stressed 
that the 
observed pattern seems to be close to the ``rank-one'' limit, in which the 
mass matrices have the ``democratic'' form of a matrix with all matrix 
elements equal to unity, which has one eigenvalue 3 and two eigenvalues 0;  
when both up and down quark mass matrices have this form, 
they are diagonalized 
by the same unitary transformation and the CKM matrix is unity.  A  
generalization of the democratic form, that is closely related to the models 
developed below, is the suggestion of Harrison and Scott [2] that 
the Hermitian 
square of the mass matrix should have the form of a circulant matrix.   
Because the underlying dynamical basis for these choices has not been 
apparent, it has not been possible to systematically extend them to  
renormalizable field theory models  
that incorporate, and relate, the observed mass and mixing 
hierarchies.  

We present in this paper models for the quark mass and flavor mixing 
matrices, based on the underlying dynamical assumption that the three 
flavor weak eigenstates are distinguished by different eigenvalues of 
a discrete chiral $Z_6$ quantum number.  The idea that a discrete chiral 
quantum number may underlie family structure was introduced originally  
by Harari and Seiberg [3], and was developed recently by 
the author [4] in a
modified form that we follow here.  Also of relevance is the remark of  
Weinberg [5] that an unbroken discrete chiral quantum number suffices 
to enforce the masslessness of fermionic states.  Extending the general  
framework of this earlier work, we postulate that
all {\it complex} fields carry a discrete chiral family quantum number.  
Since the Higgs scalars in the standard model are complex, we introduce  
one or two 
triplets of Higgs doublets that carry $Z_6$ quantum numbers, and that 
are coupled to the fermions by Yukawa couplings constructed 
so that the Lagrangian is exactly $Z_6$ invariant.  
Spontaneous symmetry breaking, in which the neutral members of the three
or six 
Higgs doublets acquire vacuum expectations, then gives the 
fermion mass matrices that form the basis for our detailed analysis.  

In addition to postulating that the Lagrangian has an exact discrete 
chiral symmetry that is spontaneously broken, we also postulate 
that there is an $S_3$ cyclic symmetry under cyclic 
permutation of the flavor eigenstates
that is explicitly but weakly broken by the Yukawa couplings and the 
Higgs self-couplings in the 
Lagrangian.  This assumption permits the analysis of our models by   
developing them in a perturbation expansion in powers of the $S_3$ cyclic 
symmetry breaking, leading, as we shall see, to qualitative features of 
the mass and mixing hierarchies that accord with observation.  An  
interplay of spontaneously broken symmetries with weakly explicitly broken 
symmetries has played a useful role in particle phenomenology in the past, 
most notably 
in understanding the consequences of chiral symmetry in quantum 
chromodynamics.  Our analysis suggests that such an interplay, in the  
context of electroweak symmetry breaking,  may also provide a basis for 
understanding features of the mass and mixing hierarchy.   

This paper is organized as follows.  In Sec. II we elaborate on the 
form of and motivation for our basic assumptions of an exact discrete 
chiral symmetry and an approximate $S_3$ cyclic permutation 
symmetry.  In Sec. III we write 
down the Lagrangians for two extensions of the standard model that 
incorporate these assumptions, the first based on a single 
three family set of Higgs 
doublets, and the second based on including an additional 
weakly coupled three family set  of Higgs 
doublets.  In Sec. IV we review the theory of circulant and retrocirculant 
matrices, in the framework of the $3 \times 3$ matrices that are needed for 
the subsequent analysis.  In Sec. V we discuss the extrema of the 
Higgs potentials in the three and six doublet models, in the limit 
of exact $S_3$ cyclic symmetry.  We work out the spectra of physical Higgs 
particles,  and show that for a wide range of parameters, the six doublet 
model leads to spontaneous violation of CP.  In a related Appendix we 
give the formulas needed for numerical minimization of the Higgs potentials  
by the conjugate gradient method.  In Sec. VI we use the extrema determined 
in Sec. V to calculate the tree approximation mass matrices.  We show that 
in the limit of exact cyclic permutation 
symmetry, the mass matrices are retrocirculants, 
corresponding to the rank one ``democratic'' form in the three doublet 
model and to a rank two generalization in the six doublet model when CP 
is spontaneously violated.  Also, in 
the limit of exact cyclic permutation 
symmetry, we characterize the Higgs decay modes, and show that 
the CKM matrix is exactly unity and that 
strangeness changing neutral currents exactly vanish.  In Sec. VII 
we formulate 
a perturbative expansion around the zeroth order approximation of exact 
$S_3$ cyclic permutation symmetry, and 
show that the mixing matrix for the first and second 
families is zeroth order in the perturbation, whereas the mixings involving 
the third family are first order in the perturbation.  In Sec. VIII we 
derive formulas for the contributions from Higgs exchange to 
the $K_L-K_S$ mass difference, which is the process most sensitive to   
strangeness changing neutral current effects.  In Sec. IX we describe the 
procedure used for making overall fits of our model, including small 
violations of cyclic permutation symmetry, to the data, give sample 
numerical results, 
and draw some conclusions from these.  In Sec. X we summarize experimental 
signatures for our model, comment on its extension to neutrino masses 
and mixings, discuss the prospects 
for coupling constant unification, and give some directions for future 
investigations.  

\section*{II.~~Basic Assumptions:  An Exact Discrete Chiral 
Symmetry and an Approximate $S_3$ Cyclic Symmetry}

In formulating our basic assumptions, we shall follow a procedure that 
has worked well in the past as an heuristic tool in particle physics. 
This is to abstract symmetry or partial symmetry assumptions from 
specific simplified field theory models, then to discard the models,   
but to retain the symmetry assumptions deduced from them as the basis 
for phenomenological calculations. Examples where this has been a productive  
method in the past include (1) the CVC and PCAC symmetries of the strong 
interactions, the algebra of currents, and the calculational methods based  
on these, and (2) the approximate $SU(3)$  flavor symmetry of the 
strong interactions.  These postulates, which had 
a somewhat {\it ad hoc} character at the time when they were first 
introduced, helped pave the way for 
the formulation of the standard model, into which they were incorporated 
in a natural way and thereby ultimately justified.  

Our aim in this paper is to apply a similar method to the problems of 
family structure and mass and mixing matrices, which to date have been 
among the most vexing puzzles of the standard model.  As an heuristic  
field theoretic model, we shall adopt a simplified composite model in 
which all matter particles (quarks, leptons, and Higgs fields -- everything 
other than the gauge fields) are composites of a single fermion field 
$\chi$.  As observed by Harari and Seiberg [3] and Weinberg [5], 
in a gauge theory for $\chi$ the instanton determinant that breaks 
global $U(1)$ invariance leaves unbroken a discrete $Z_{2K}$ chiral subgroup, 
with $K$ determined by the index of the representation of the gauge group 
under which $\chi$ transforms.  Harari and Seiberg propose, moreover,  
that this naturally occurring discrete chiral subgroup provides the 
quantum number that distinguishes between the various families.  Since 
it is now clear that there are exactly three light families, we shall 
assume henceforth in applying this idea that $K=3$, so that we start 
from the assumption that the 
fundamental Lagrangian, as augmented by the instanton-induced potential, 
is invariant under the simultaneous transformations 
\begin{mathletters}
\label{allequations}
\begin{equation}
\chi_L \to \chi_L \exp(2\pi i/6)~,~~~\chi_R \to \chi_R \exp(-2\pi i/6)~~~
\label{equationa}
\end{equation}
of the fundamental fermion fields $\chi$.  
The fields in the low energy 
effective Lagrangian are in general nonlinear functionals of the fundamental 
fields.  Fermionic effective fields must be odd monomials in the fundamental 
fields, and so can come in three varieties $\psi_n$ with the discrete 
chiral transformation law
\begin{equation}
\psi_{nL} \to \psi_{nL} \exp((2n+1)2\pi i/6)~,~~~
\psi_{nR} \to \psi_{nR} \exp(-(2n+1)2\pi i/6)~,~~~
n=1,2,3~~~,
\label{equationb}
\end{equation}
while complex bosonic effective fields must be even monomials in 
the fundamental fermion fields, and so can 
also come in three varieties $\phi_n$ 
with the discrete chiral transformation law 
\begin{equation}
\phi_n \to \phi_n \exp(2n 2\pi i/6)~,~~~n=1,2,3~~~.\label{equationc}
\end{equation}
\end{mathletters}
Introducing the cube roots of 
unity $\omega$ and $\overline \omega$, 
\begin{mathletters}
\label{allequations}
\begin{equation}
\omega=\exp(2\pi i/3)=-{1 \over 2} +{ \surd 3 \over 2}i,~~~
\overline \omega=\exp(-2 \pi i/3)=-{1 \over 2} -{ \surd 3 \over 2}i~~~,
\label{equationa}
\end{equation}
that obey the relations 
\begin{equation}
\overline  \omega=\omega^*=\omega^2~,~~~1+\omega+\overline \omega=0~~~,
\label{equationb}
\end{equation}
\end{mathletters}
the transformation laws of Eqs.~(1a-c) take the form
\begin{eqnarray}
\chi_L \to&& \chi_L \omega^{1\over 2}~,
~~~\chi_R \to \chi_R \overline \omega^{1\over 2}~~, \nonumber\\
\psi_{nL} \to&& \psi_{nL} \omega^{n+{1\over 2}}~,~~~
\psi_{nR} \to \psi_{nR} {\overline \omega}^{n+{1\over
2}}~,~~~n=1,2,3,\nonumber \\
\phi_n \to&& \phi_n \omega^n~,~~~n=1,2,3~~~. 
\end{eqnarray}

Gauge fields are real fields, and since the phase in Eq.~(1c) never takes 
the value $-1$ for any $n$,   
the gauge fields in a $Z_6$ model necessarily come in only one variety  
transforming with phase 
unity under discrete chiral transformations.   
Thus the minimal $Z_6$ invariant extension of the standard  
model consists of a triplicated set of fermions, and a triplicated set of 
Higgs doublets, obeying the transformation laws of Eqs.~(1b) and (1c) 
respectively, together with the usual gauge bosons,  with the Lagrangian 
constructed to be $Z_6$ invariant.  

As we shall see in Sec.~III below, the assumption of an unbroken discrete 
chiral symmetry still leaves many parameters in the Lagrangian, and it 
is desirable to look for a further exact or approximate symmetry to 
impose.  The natural candidate is $S_3$ cyclic permutation symmetry, under 
simultaneous cyclic permutation of the $n=1,2,3$ discrete 
chiral components of 
the fermion and Higgs boson fields.  If the discrete chiral components were 
physically identical, one would expect this $S_3$ cyclic 
symmetry to be exact. 
However, in the composite picture from which we are abstracting our model,  
the discrete chiral components differ physically by the addition of 
fermion-antifermion pairs coupled as Lorentz scalars, and so the internal 
wave functions of the discrete chiral components are different.  Thus 
the best we might hope for is an approximate, weakly broken, $S_3$ cyclic 
permutation symmetry, and this will be assumed as the second 
ingredient of our model. 

By abstracting our two fundamental assumptions from a schematic composite 
model, we gain some assurance that they are consistent with each other   
and at least physically plausible.   However, we do not attach great 
significance to the particular model from which they were inferred; it is 
entirely possible that the same assumptions can emerge from other dynamical 
frameworks.  We shall henceforth avoid further discussion of 
underlying models, and focus on exploring the consequences of our 
assumptions within the standard framework of low energy renormalizable 
effective action phenomenology.  

\bigskip
\section*{III.~~Discrete Chiral Invariant Extensions of the Standard Model}

We proceed now to write down discrete chiral invariant extensions of 
the Lagrangian density for the standard model, following the notation of the 
text of Mohapatra [6].  In the following, each quark or lepton field 
is implicitly a column vector formed 
from the three discrete chiral components obeying the transformation laws 
of Eq.~(3), with the $n=1$ index at 
the top of the column vector and the $n=3$ index at the bottom. For the    
Higgs scalar fields, the discrete chiral subscript $n$ will be indicated 
explicitly.  
We shall be interested   
in two models, the first containing a single discrete chiral triplet of Higgs 
doublets $\phi$, the second containing two discrete chiral triplets of 
Higgs doublets, denoted respectively by $\phi$ and $\eta$.   We shall write 
all formulas for the case of the six Higgs doublet model; the simpler  
three doublet model is obtained by setting all fields $\eta$ to zero.  

The total Lagrangian density ${\cal L}$ consists of kinetic terms for 
the gauge, Higgs, and fermionic fields,  together with Yukawa 
couplings of the Higgs fields to  the fermions and the Higgs 
self-interaction potential.    
Writing   
\begin{mathletters}
\label{allequations}
\begin{eqnarray}
{\cal L}=&&{\cal L}_{\rm gauge~kinetic} + {\cal L}_{\rm Higgs~kinetic} 
+{\cal L}_{\rm fermion~kinetic}\nonumber\\
+&&{\cal L}_{\rm Yukawa} +{\cal L}_{\rm
Higgs~potential}~~~,\label{equationa}
\end{eqnarray}
the gauge kinetic terms have the usual form 
\begin{eqnarray}
{\cal L}_{\rm gauge~kinetic}=&&-{1\over 4}\vec W_{\mu\nu} 
\cdot \vec W_{\mu\nu}
-{1\over 4}B_{\mu\nu}B_{\mu\nu}~~~,\nonumber\\
\vec W_{\mu\nu}=&&\partial_{\mu}\vec W_{\nu}-\partial_{\nu}\vec W_{\mu}
+g\vec W_{\mu} \times \vec W_{\nu}~,~~
~B_{\mu \nu}=\partial_{\mu}B_{\nu}-\partial_{\nu}B_{\mu}~~~,
\label{equationb}
\end{eqnarray}
and so also do the fermion kinetic terms (with $Q_L$ and $\psi_L$ 
respectively the left-handed quark and lepton doublets, and $\vec \tau$ the 
weak isospin Pauli matrices that act on them), 
\begin{eqnarray}
{\cal L}_{\rm fermion~kinetic}=&&-\overline Q_L \gamma_{\mu}
(\partial_{\mu} -{ig \over 2} \vec \tau \cdot \vec W_{\mu}
-{i g^{\prime} \over 6} B_{\mu} ) Q_L  \nonumber\\
&&-\overline{\psi}_L \gamma_{\mu} (\partial_{\mu} -{ig \over 2} 
\vec \tau \cdot \vec W_{\mu}+{i g^{\prime} \over 6} B_{\mu} )\psi_L \nonumber\\
&&-\overline e_R \gamma_{\mu}(\partial_{\mu}+ i g^{\prime}B_{\mu})e_R
-\overline \nu_R \gamma_{\mu} \partial_{\mu} \nu_R \nonumber\\
&&-\overline u_R \gamma_{\mu} 
(\partial_{\mu}-{2i g^{\prime}\over 3} B_{\mu})u_R
-\overline d_R \gamma_{\mu}(\partial_{\mu}+{i g^{\prime} \over 3} B_{\mu} )
d_R ~~~.
\label{equationc}
\end{eqnarray}
The Higgs kinetic energy is simply a sum over kinetic 
terms of the standard form 
for the discrete chiral components of the scalars $\phi$ and $\eta$  (each  
of which is, as usual, a weak isospin doublet), 
\begin{eqnarray}
{\cal L}_{\rm Higgs~kinetic}
=&&-\sum_{n=1,2,3}|\partial_{\mu}\phi_n
-{ig\over 2}\vec \tau \cdot \vec W_{\mu} \phi_n
-{i g^{\prime}\over 2} B_{\mu} \phi_n|^2  \nonumber\\
&&-\sum_{n=1,2,3}|\partial_{\mu}\eta_n
-{ig\over 2}\vec \tau \cdot \vec W_{\mu} \eta_n
-{i g^{\prime}\over 2} B_{\mu} \eta_n|^2~~~. 
\label{equationd}
\end{eqnarray}
\end{mathletters}

It is only in the Yukawa couplings and the Higgs potential that invariance 
under discrete chiral transformations plays a nontrivial role.   
Letting $\tilde \phi_n$ and $\tilde \eta_n$ denote the CP conjugates 
of the Higgs fields, 
\begin{mathletters}
\label{allequations}
\begin{eqnarray}
\tilde \phi_n=&&(CP)^{-1} \phi_n CP=i \tau_2 \phi_n^*~~~,\nonumber\\
\tilde \eta_n=&&(CP)^{-1} \eta_n CP=i \tau_2 \eta_n^*~~~,
\label{equationa}
\end{eqnarray}
the Yukawa Lagrangian takes the form
\begin{equation}
{\cal L}_{\rm Yukawa} =
\overline Q_L \Phi^d d_R + \overline Q_L \Phi^u u_R
+\overline \psi_L \Phi^e e_R + \overline \psi_L \Phi^{\nu} \nu_R~~~
+{\rm adjoint},
\label{equationb}
\end{equation}
\end{mathletters}
where $\Phi^f,~f=d,u,e,\nu$ is a $3\times 3$ matrix acting on the 
discrete chiral column vector structure, and where we have allowed for 
the possibility of nonzero Dirac neutrino masses by including a right-handed 
neutrino.  The matrices $\Phi^f$ must be constructed so that Eq.~(5b) 
is invariant under simultaneous discrete chiral transformations of the 
fermion and Higgs fields.  Referring to Eq.~(3), it is easy to see that 
this dictates the structure, 
\begin{mathletters}
\label{allequations}
\begin{eqnarray}
\Phi^f=&&g_{\phi}^f(P_{\phi 1}^f \phi_1 + P_{\phi 2}^f \phi_2 
+ P_{\phi 3}^f \phi_3) + g_{\eta}^f (P_{\eta 1}^f \eta_1
+P_{\eta 2}^f \eta_2 + P_{\eta 3}^f \eta_3),~f=d,e~~~,\nonumber\\
\Phi^f=&&g_{\phi}^f(P_{\phi 1}^f \tilde \phi_2 + P_{\phi 2}^f \tilde \phi_1 
+ P_{\phi 3}^f \tilde \phi_3) + g_{\eta}^f (P_{\eta 1}^f \tilde \eta_2
+P_{\eta 2}^f \tilde \eta_1 + P_{\eta 3}^f \tilde \eta_3),~f=u,\nu~~~,
\label{equationa}
\end{eqnarray}
with the $3\times 3$ matrices $P_{\xi n}^f$ given, 
for all flavors $f=u,d,e,\nu$ and 
for $\xi=\phi,\eta$, by 
\begin{eqnarray}
P_{\xi 1}^f=&&
  \pmatrix{ 0 & 1+\beta_{\xi 12}^f & 0\cr   
           1+\beta_{\xi 21}^f & 0 & 0 \cr
           0 & 0 & 1+\beta_{\xi 33}^f \cr} ~~~,\nonumber\\
&&\nonumber\\
P_{\xi 2}^f=&&
  \pmatrix{ 0 & 0 &1+\beta_{\xi 13}^f \cr   
           0& 1+\beta_{\xi 22}^f & 0  \cr
            1+\beta_{\xi 31}^f& 0 & 0 \cr} ~~~,\nonumber\\
&&\nonumber\\
P_{\xi 3}^f=&&
  \pmatrix{ 1+\beta_{\xi 11}^f & 0 & 0\cr   
            0 & 0& 1+\beta_{\xi 23}^f  \cr
           0 &  1+\beta_{\xi 32}^f& 0 \cr} ~~~.
\label{equationb}
\end{eqnarray}
To uniquely specify the Yukawa couplings $g_{\xi}^f$, we require that 
the parameters $\beta_{\xi mn}^f$ sum to zero, 
\begin{equation}
\sum_{mn} \beta_{\xi mn}^f=0~~~. 
\label{equationc}
\end{equation}
\end{mathletters}
When there is exact $S_3$ cyclic permutation symmetry 
the $\beta$'s all vanish,  
and thus the case of approximate $S_3$ cyclic symmetry is parameterized 
by $\beta$'s that are all small compared to unity. In a CP conserving 
theory all of the coupling constants $g_{\phi,\eta}^f$ and all of the 
$\beta$'s are real; when CP conservation is not imposed, these parameters 
can be complex.  

We turn finally to the Higgs potential, which we separate into four 
terms as follows,
\begin{mathletters}
\label{allequations}
\begin{equation}
{\cal L}_{\rm Higgs~potential}= V_{\phi}+V_{\eta} 
+V_1(\phi,\eta) + V_2(\phi,\eta)~~~,
\label{equationa}
\end{equation}
with (for $\xi=\phi,\eta$)
\begin{eqnarray}
V_{\xi}=&&\sum_{n=1}^3 V_{\xi n}~~~,  \nonumber\\
V_{\xi n}=&&\lambda_{\xi n} (\xi_n^{\dagger} \xi_n -v_{\xi n}^2)^2
-\mu_{1\xi n} \xi_n^{\dagger}\xi_n \xi_{n+1}^{\dagger}\xi_{n+1}
-\mu_{2\xi n}|\xi_n^{\dagger}\xi_{n+1}|^2
-\alpha_{\xi n}{\rm Re} \exp(i \psi_{\xi n}) \xi_n^{\dagger} \xi_{n+1}
\xi_n^{\dagger} \xi_{n-1}~,
\label{equationb}
\end{eqnarray}
where the coefficients in Eq.~(7b) are real (by hermiticity) and where 
the parameter $\psi_{\xi n}$ is zero (modulo $\pi$) 
when CP conservation is imposed. For the potential terms 
that couple the $\phi$ and $\eta$ Higgs fields, 
we have in the CP conserving case
\begin{eqnarray}
V_1(\phi,\eta)=&&\sum_{m,n=1}^3  
(C_{1 mn} \phi_m^{\dagger} \phi_m \eta_n^{\dagger}\eta_n
+C_{2 mn} {\rm Re} \phi_m^{\dagger}\eta_m \eta_n^{\dagger}\phi_n \nonumber\\
+&&C_{3 mn} {\rm Re}\phi_m^{\dagger}\phi_{m+1}\eta_n^{\dagger}\eta_{n-1}
+C_{4 mn} {\rm Re} \eta_m^{\dagger} \eta_{m+1} 
\phi_n^{\dagger} \phi_{n-1} \nonumber\\
+&&C_{5 mn} {\rm Re} \phi_m^{\dagger} \eta_{m+1} \eta_n^{\dagger} \phi_{n-1} 
+C_{6 mn} {\rm Re} \eta_m^{\dagger} \phi_{m+1} \phi_n^{\dagger} \eta_{n-1} )  
~~~,\nonumber\\
V_2(\phi,\eta)=&&\sum_n  \gamma_n {\rm Re} \phi_n^{\dagger} \eta_n \nonumber\\
+&&\sum_{m,n=1}^3   
(C_{7 mn} {\rm Re} \phi_m^{\dagger} \phi_{m+1} \phi_n^{\dagger}\eta_{n-1}
+C_{8 mn} {\rm Re} \phi_m^{\dagger}\phi_{m+1}
\eta_n^{\dagger}\phi_{n-1} \nonumber\\ 
+&&C_{9 mn} {\rm Re}\eta_m^{\dagger}\eta_{m+1}\phi_n^{\dagger}\eta_{n-1}
+C_{10 mn} {\rm Re} \eta_m^{\dagger} \eta_{m+1} \eta_n^{\dagger} 
\phi_{n-1} \nonumber\\
+&&C_{11 mn} {\rm Re} \phi_m^{\dagger} \eta_{m+1} \phi_n^{\dagger} \eta_{n-1} 
+C_{12 mn} {\rm Re} \eta_m^{\dagger} \phi_{m+1} \eta_n^{\dagger} \phi_{n-1} )  
~~~,
\label{equationc}
\end{eqnarray}
\end{mathletters}
with all constants real (again by hermiticity).  The terms $V_1$ are those 
invariant under independent rephasings 
$\phi_n \to \exp(i \theta_{\phi}) \phi_n$  and  $\eta_n \to 
\exp(i \theta_{\eta}) \eta_n$ of the two Higgs discrete chiral triplets, 
while the terms $V_2$ are only invariant under this 
phase transformation when restricted so that  
$\theta_{\phi}=\theta_{\eta}$.   
When CP is not conserved, an independent phase can be inserted inside 
each real part Re in the above expressions, in analogy with the 
construction of the final term of Eq.~(7b).  
When there is $S_3$ cyclic permutation symmetry, the constants with a single 
discrete chiral subscript $n$ are independent of that subscript, while 
the constants with a double subscript $mn$ obey the cyclic condition 
$C_{\ell mn}=C_{\ell m+1\,n+1},~~~\ell=1,...,12~$.

This rather complicated Higgs potential completes the specification of 
our model, the tree approximation to which will be analyzed in detail  
in the sections that follow. 

\section*{IV.~~Retrocirculant and Circulant Matrices}  
 
Before proceeding further, we pause to review the theory of circulant  
and retrocirculant matrices in the $3 \times 3$ case relevant for what 
follows. For a compact summary of general results see Marcus [7] and 
Hamburger and Grimshaw [7], and    
for a detailed exposition see Davis [8].  A matrix 
\begin{mathletters}
\label{allequations}
\begin{equation}
{\rm Circ}_{\rightarrow}(a,b,c) \equiv
  \pmatrix{a & b& c \cr
           c &a & b \cr
           b &c &a \cr}~~~~, 
\label{equationa}
\end{equation}
is called a {\it circulant}, while a matrix 
\begin{equation}
{\rm Circ}_{\leftarrow}(a,b,c) \equiv
  \pmatrix{a & b& c \cr
           b &c & a \cr
           c &a &b \cr}~~~~, 
\label{equationb}
\end{equation}
\end{mathletters}
is called a {\it reverse circulant} or  {\it retrocirculant}.  [Clearly, 
a retrocirculant is always a symmetric matrix, and so 
${\rm Circ}_{\leftarrow}(a,b,c)={\rm Circ}_{\leftarrow}(a,b,c)^T$, and 
${\rm Circ}_{\leftarrow}(a,b,c)^{\dagger}={\rm Circ}_{\leftarrow}(a,b,c)^*$.]
Two properties of these matrices are used in what follows.  The first is 
that the Hermitian square of a retrocirculant is a circulant, 
\begin{eqnarray}
{\rm Circ}_{\leftarrow}(a,b,c) {\rm Circ}_{\leftarrow}(a,b,c)^{\dagger}
=&&{\rm Circ}_{\rightarrow}(|a|^2+|b|^2+|c|^2,ab^*+bc^*+ca^*,ac^*+ba^*+cb^*)
~~~,\nonumber\\
{\rm Circ}_{\leftarrow}(a,b,c)^{\dagger} {\rm Circ}_{\leftarrow}(a,b,c)
=&&{\rm Circ}_{\rightarrow}(|a|^2+|b|^2+|c|^2,a^*b+b^*c+c^*a,a^*c+b^*a+c^*b)
~~~.
\end{eqnarray}
The second is that any retrocirculant with arbitrary complex 
$a,b,c$ is diagonalized by 
transformation from the left and right by unitary matrices $U_L$, 
$U_R=U_L^*$,  
that are independent of the values of $a,b,c$.  Explicitly, setting 
\begin{mathletters}
\label{allequations}
\begin{eqnarray}
U_L^{~}=&&{1 \over \surd 3} \pmatrix{ 1 & \overline \omega& \omega \cr
                                  1 & \omega & \overline \omega \cr
                                  1 & 1      & 1
\cr}~~~,\nonumber\\ 
&&\nonumber\\
U_R^{~}=&&{1 \over \surd 3} \pmatrix{ 1 &  \omega& \overline \omega \cr
                                  1 & \overline \omega &  \omega \cr
                                  1 & 1      & 1
\cr}~~~,\nonumber\\ 
&&\nonumber\\
U_R^{\dagger}=&&{1 \over \surd 3} \pmatrix{ 1 &  1& 1 \cr
                                  \overline \omega & \omega & 1 \cr
                                  \omega  & \overline \omega & 1
\cr}~~~,
\label{equationa}
\end{eqnarray}
a simple calculation shows that                                   
\begin{equation}
U_L {\rm Circ}_{\leftarrow}(a,b,c) U_R^{\dagger}=
\pmatrix{a+\overline \omega b + \omega c & 0 & 0 \cr
0 & a + \omega b + \overline \omega c & 0 \cr
0 & 0&  a + b + c \cr}~~~. 
\label{equationb}
\end{equation}
\end{mathletters}
An elementary corollary of these statements is that any Hermitian 
circulant matrix $H_{\rightarrow}$ is diagonalized by the 
unitary transformation $U_L H_{\rightarrow} U_L^{\dagger}$ using the 
unitary matrix $U_L$ of Eq.~(10b).  

The relevance of these results to what follows is that  
in the limit of 
$S_3$ cyclic permutation symmetry, we shall find that 
the fermion mass matrices in both the three and 
six doublet models are retrocirculants, and so are diagonalized by the 
universal bi-unitary transformation of Eq.~(10b).  By Eq.~(9),    
the Hermitian squares of the fermion mass matrices in the   
approximation of cyclic permutation symmetry are therefore circulants, 
as suggested by Harrison and Scott [2].   We shall further find, in analyzing 
the Higgs sector in the case of cyclic permutation symmetry, that the Higgs 
mass matrices are also circulants, making it easy to diagonalize them    
explicitly.  

\bigskip
\section*{V.~~Structure of the Higgs Sector}  

We turn now to an analysis of the properties of the discrete chiral 
invariant Higgs potential 
of Eqs.~(7a-c).  We shall assume CP invariance and exact $S_3$ cyclic 
permutation 
symmetry; when needed, we can take into account small deviations from 
these assumptions by adding  perturbations to the locations of the 
Higgs minima.  We begin our discussion with the three Higgs model, in 
which only the discrete chiral triplet $\phi$ is present.  Omitting the 
subscript $\phi$ on the coefficients, we have 
\begin{mathletters}
\label{allequations}
\begin{eqnarray}
{\cal L}_{\rm Higgs~potential}   
=&&\lambda \sum_{n=1}^3 (\phi_n^{\dagger} \phi_n-v^2)^2
-\mu_1 \sum_{n=1}^3 \phi_n^{\dagger}\phi_n \phi_{n+1}^{\dagger}\phi_{n+1}
-\mu_2 \sum_{n=1}^3 |\phi_n^{\dagger} \phi_{n+1}|^2 \nonumber\\
-&&\alpha \sum_{n=1}^3 {\rm Re} \phi_n^{\dagger}\phi_{n+1}
\phi_n^{\dagger}\phi_{n-1}~.
\label{equationa}
\end{eqnarray}
Necessary conditions for this potential to be bounded below are
evidently
\begin{equation}
\lambda>0,~~~\lambda-\mu_1-\mu_2-\alpha>0 ~~~.
\label{equationb}
\end{equation}
\end{mathletters}
Imposing the condition 
\begin{mathletters}
\label{allequations}
\begin{equation}
\mu_2+\alpha>0 ~~~
\label{equationa}
\end{equation}
insures that the Higgs potential is minimized when the three
doublets all have the same form
\begin{equation}
\phi_n=\pmatrix{ 0 \cr
                  \Omega_n\cr} ~~~,
\label{equationb}
\end{equation}
for a suitable choice of $SU(2)$ gauge, 
with the consequence that one electroweak gluon (the photon) remains 
massless.  Imposing the additional condition 
\begin{equation}
\alpha>0~~~
\label{equationc}
\end{equation}
\end{mathletters}
then forces the complex phases of the three expectations $\Omega_n$ to 
be equal (up to discrete chiral rephasings) 
at the minimum of the potential; by a choice of $U(1)$ gauge 
the overall common 
phase can be rotated to zero, and so the potential of Eq.~(11a) is 
minimized at 
\begin{mathletters}
\label{allequations}
\begin{equation}
\Omega_1=\Omega_2=\Omega_3=\Omega~~~,
\label{equationa}
\end{equation}
with $\Omega$ given by 
\begin{equation}
\Omega^2={\lambda v^2 \over 
\lambda - \mu_1 -\mu_2 -\alpha}~~~.
\label{equationb}
\end{equation}
This minimum is not unique; because the potential of Eq.~(11a) is 
invariant under the discrete chiral transformation of Eq.~(3), equivalent 
minima are located at 
\begin{equation}
\Omega_n=\omega_n\Omega~,~~~n=1,2,3~~~,
\label{equationc}
\end{equation}
\end{mathletters}
with $\omega_{1,2,3}$ any three distinct cube roots of unity,  
which can always be obtained by permutation from the set 
$\overline \omega, \omega, 1$.  Despite the appearance of complex 
phases in Eq.~(13c), there is no breakdown of CP invariance, because these 
phases can always be eliminated by the discrete chiral transformation that 
returns to the minimum of Eq.~(13a).  

We note that although the potential of Eq.~(11a) is similar in form to that   
studied by Bigi and Sanda [9], they choose $\alpha<0$, in which 
case there are nontrivial relative phases (that are not just discrete 
chiral rephasings) between the three 
expectations $\Omega_{1,2,3}$ at the potential minimum,  
and CP is spontaneously broken.   This 
case is not useful for our model building because numerical analysis 
shows that it leads to a mass matrix with one heavy family, and two 
other lighter families of {\it equal} mass.   We shall make use of the 
possibility [10] of CP violation in multi-Higgs systems only in the context 
of the six doublet model, to be discussed shortly.   

To complete our discussion of the three doublet model, we must determine 
the Higgs masses.  Expanding around the minimum of Eqs.~(13a, b) to 
second order by 
substituting 
\begin{mathletters}
\label{allequations}
\begin{equation}
\phi_n=\pmatrix{{1\over \surd 2} \delta_n \cr
                  \Omega+{1 \over \surd 2}\epsilon_n\cr} ~~~
\label{equationa}
\end{equation}
into Eq.~(11a), we find 
\begin{eqnarray}
{\cal L}_{\rm Higgs~potential}=&&V_0 + V_{2\delta} +V_{2\epsilon}~~~,
\nonumber\\ 
V_{2\delta}=&&\sum_{m,n=1}^3{1\over 2}\delta_m^*B_{mn} \delta_n~~~,
\nonumber\\ 
V_{2\epsilon}=&&\sum_{m,n=1}^3{1\over 2}[\epsilon_m^* A_{mn} \epsilon_n
+\epsilon_m^* D_{mn} \epsilon_n^*+\epsilon_m D_{mn} \epsilon_n]~~~.
\label{equationb}
\end{eqnarray}
\end{mathletters}
A simple calculation shows that the matrices $A,B,D$ are all circulants  
of the form
\begin{mathletters}
\label{allequations}
\begin{eqnarray}
A=&&{\rm Circ}_{\rightarrow}(a^A,b^A,b^A)~~~, \nonumber\\
B=&&{\rm Circ}_{\rightarrow}(a^B,b^B,b^B)~~~, \nonumber\\
D=&&{\rm Circ}_{\rightarrow}(a^D,b^D,b^D)~~~,
\label{equationa}
\end{eqnarray}
with $a^{A,B,D}$ and $b^{A,B,D}$ given in terms of the Lagrangian parameters  
by
\begin{eqnarray}
a^A=&&(2\lambda-\mu_1-\mu_2)2\Omega^2-2\lambda v^2~,~~~
b^A=-(\mu_1+\mu_2+2\alpha)\Omega^2~~~,  \nonumber\\
a^B=&&2(\lambda-\mu_1)\Omega^2-2 \lambda v^2~,~~~
b^B=-(\mu_2+\alpha)\Omega^2~~~,\nonumber\\
a^D=&&(\lambda-{1\over 2}\alpha)\Omega^2~,~~~
b^D=-{1\over 4}(2\mu_1 + 2\mu_2 + \alpha)\Omega^2~~~.
\label{equationb}
\end{eqnarray}
\end{mathletters}
Since these matrices are all diagonalized by transformations  
based on the cube roots of unity, it is useful to introduce new bases  
defined as follows,  
\begin{mathletters}
\label{allequations}
\begin{eqnarray}
\pmatrix{\phi_1 \cr \phi_2 \cr \phi_3 \cr}=&&
W \pmatrix{\phi^{(1)} \cr\phi^{(2)} \cr \phi^{(3)} \cr}~,~~~
\pmatrix{\phi^{(1)} \cr\phi^{(2)} \cr \phi^{(3)} \cr}=
W^{-1}\pmatrix{\phi_1 \cr \phi_2 \cr \phi_3 \cr}~~~,\nonumber\\
&&\nonumber\\
\pmatrix{\delta_1 \cr \delta_2 \cr \delta_3 \cr}=&&
W \pmatrix{\delta^{(1)} \cr\delta^{(2)} \cr \delta^{(3)} \cr}~,~~~
\pmatrix{\delta^{(1)} \cr\delta^{(2)} \cr \delta^{(3)} \cr}=
W^{-1}\pmatrix{\delta_1 \cr \delta_2 \cr \delta_3 \cr}~~~,\nonumber\\
&&\nonumber\\
\pmatrix{\epsilon_1 \cr \epsilon_2 \cr \epsilon_3 \cr}=&&
W \pmatrix{\epsilon^{(1)} \cr\epsilon^{(2)} \cr \epsilon^{(3)} \cr}~,~~~
\pmatrix{\epsilon^{(1)} \cr\epsilon^{(2)} \cr \epsilon^{(3)} \cr}=
W^{-1}\pmatrix{\epsilon_1 \cr \epsilon_2 \cr \epsilon_3 \cr}~~~,
\label{eqationa}
\end{eqnarray}
with 
\begin{eqnarray}
W=W^T=&&
{1\over \surd 3}\pmatrix{ \omega & \overline \omega & 1 \cr
\overline \omega & \omega & 1 \cr 1& 1& 1\cr }~,~~~
W^{-1}=W^{\dagger}=W^*=
{1\over \surd 3}\pmatrix{ \overline \omega &  \omega & 1 \cr
 \omega & \overline \omega & 1 \cr 1& 1& 1\cr }~,\nonumber\\
&&\nonumber\\
W^{\dagger} {\rm Circ}_{\rightarrow}(a,b,c) W=&&
\pmatrix{a+\omega b+ \overline \omega c &0&0\cr
0& a+\overline \omega b +\omega c &0 \cr
0& 0& a+b+c \cr}~,~~~\nonumber\\
&&\nonumber\\
W{\rm Circ}_{\rightarrow}(a,b,c) W=&&
\pmatrix{0 &a+\overline \omega b+  \omega c &0\cr
 a+ \omega b +\overline \omega c &0&0 \cr
0& 0& a+b+c \cr}~~~.
\label{equationb}
\end{eqnarray}
In terms of the new bases, Eq.~(14a) becomes
\begin{equation}
\phi^{(n)}=\pmatrix{ {1\over \surd 2} \delta^{(n)}\cr 
                  {1 \over \surd 2}\epsilon^{(n)}\cr},~~~n=1,2,
\label{equationc}
\end{equation}
and 
\begin{equation}
\phi^{(3)}=\pmatrix{{1\over \surd 2} \delta^{(3)}\cr 
    \surd 3 \Omega+{1 \over \surd 2}\epsilon^{(3)}\cr}~~~.
\label{equationd}
\end{equation}
\end{mathletters}

Substituting Eq.~(16b) into both Eq.~(14b) and the Higgs kinetic energy, 
and using Eq.~(13b), 
we find for the terms quadratic in $\delta_n$
\begin{mathletters}
\label{allequations}
\begin{eqnarray}
&&-\sum_{n=1,2,3}{1\over 2}|\partial_{\mu} \delta_n|^2 + V_{2\delta}\nonumber\\
&&=-\sum_{n=1,2,3}{1\over 2}|\partial_{\mu} \delta^{(n)}|^2         
+(a^B+2b^B){1\over  2}|\delta^{(3)}|^2 +(a^B-b^B){1\over 2}
(|\delta^{(1)}|^2+|\delta^{(2)}|^2) \nonumber\\
&&=-\sum_{n=1,2,3}{1\over 2}|\partial_{\mu} \delta^{(n)}|^2      
+{3 \over 2} (\mu_2 + \alpha) \Omega^2 
(|\delta^{(1)}|^2+|\delta^{(2)}|^2) ~~~.
\label{equationa}
\end{eqnarray}
From Eq.~(17a) we see that $\delta^{(3)}$ 
is a charged massless Goldstone 
boson (which is absorbed by the Higgs mechanism into the longitudinal 
parts of the charged intermediate bosons), while $\delta^{(1,2)}$ are 
two charged Higgs boson fields  
(each containing a positive and a negative charge 
state) , with mass squared $3(\mu_2+\alpha)\Omega^2$.  
Similarly, we find for the terms quadratic in $\epsilon_n$
\begin{eqnarray}
&&-\sum_{n=1,2,3}{1\over 2}|\partial_{\mu} \epsilon_n|^2 +
V_{2\epsilon}  \nonumber\\ 
&&=-\sum_{n=1,2,3}{1\over 2}|\partial_{\mu} \epsilon^{(n)}|^2        
+(a^A+2b^A){1\over  2}|\epsilon^{(3)}|^2 +(a^A-b^A){1\over 2}
(|\epsilon^{(1)}|^2+|\epsilon^{(2)}|^2)\nonumber\\
&&+(a^D+2b^D){1 \over 2}[(\epsilon^{(3)})^2 + (\epsilon^{(3)*})^2]
+(a^D-b^D)(\epsilon^{(1)}\epsilon^{(2)}+\epsilon^{(1)*}\epsilon^{(2)*})~~~.
\label{equationb}
\end{eqnarray}
Defining new
linear combinations $\epsilon^{(\pm)}$ by 
\begin{equation}
\epsilon^{(\pm)}={1 \over \surd 2}(\epsilon^{(1)}
\pm \epsilon^{(2)})~~~,
\label{equationc}
\end{equation}
and splitting $\epsilon^{(3)},~\epsilon^{(\pm)}$ into 
real and imaginary parts, 
$\epsilon^{(3,\pm)}=\epsilon^{(3,\pm)}_R+i\epsilon^{(3,\pm)}_I$,  
Eq.~(17b) takes the form
\begin{eqnarray}
&&-\sum_{n=1,2,3}{1\over 2}|\partial_{\mu} \epsilon_n|^2 +
V_{2\epsilon} \nonumber\\ 
&&=-{1 \over 2}[(\partial_{\mu} \epsilon_R^{(3)})^2                       
+(\partial_{\mu} \epsilon_I^{(3)})^2 
+ (\partial_{\mu} \epsilon_R^{(+)})^2+(\partial_{\mu} \epsilon_I^{(+)})^2  
+ (\partial_{\mu} \epsilon_R^{(-)})^2+(\partial_{\mu} \epsilon_I^{(-)})^2]  
+4\lambda v^2 {1\over 2} (\epsilon_R^{(3)})^2 \nonumber\\
&&+(4 \lambda+2 \mu_1 + 2 \mu_2 +{7 \over 2} \alpha)\Omega^2 
{1\over 2}[(\epsilon_R^{(+)})^2 + (\epsilon_I^{(-)})^2 ]    
+{9 \over 2} \alpha \Omega^2     
{1\over 2}[(\epsilon_R^{(-)})^2 + (\epsilon_I^{(+)})^2 ]  ~~~.
\label{equationd}
\end{eqnarray}
\end{mathletters}
We see that $\epsilon_I^{(3)}$ is a neutral 
massless Goldstone boson (which is 
absorbed by the Higgs mechanism into the longitudinal part of the neutral 
intermediate boson), while $\epsilon_R^{(3)}$, both $\epsilon_R^{(+)}$ 
and $\epsilon_I^{(-)}$, and both $\epsilon_R^{(-)}$ and $\epsilon_I^{(+)}$, 
are neutral Higgs states, with respective 
squared masses $4 \lambda v^2$, $(4 \lambda+2 \mu_1 + 2 \mu_2 
+{7 \over 2} \alpha)\Omega^2$, and ${9 \over 2} \alpha \Omega^2$. 
Thus,  
the twelve states contained in the original 
triplet of Higgs doublets are accounted for as one neutral and 
two charged Goldstone modes, four charged Higgs bosons, and 
five neutral Higgs bosons.   
This information is summarized in Table I, which also gives the 
couplings of the Higgs bosons to fermions worked out in Sec. VI.  

We turn next to the properties of the Higgs sector of the six doublet 
model.  Although we shall focus here on analytic results, we have also 
made numerical studies of the minima of the six (and three) doublet 
potentials, using the formulas and method given in Appendix A. 
Let us begin by assuming that the potentials $V_1(\phi,\eta)$ and   
$V_2(\phi,\eta)$ of Eq.~(7c), that couple the $\phi$ and $\eta$ Higgs 
discrete chiral triplets, are very small.  Then the minima of the Higgs
potential are obtained by examining the degenerate minima of $V_{\phi}$ 
and $V_{\eta}$, as analyzed in the three Higgs discussion above, and 
selecting those for which $V_1 +V_2$ is smallest.  By a simultaneous 
$Z_6$ rephasing of $\phi$ and $\eta$, we can always make the minimizing 
values of $\phi$ have the form of Eqs.~(12b) and (13a), with $\Omega$ and 
the coefficients 
$\lambda,~v,~\alpha,~\mu_1,~\mu_2$  in Eq.~(13b) 
now carrying the subscript $\phi$ to differentiate them from the similar 
formulas that hold for the Higgs field $\eta$.  There are now two distinct 
possibilities, depending on the values of the coefficients in $V_1$ and 
$V_2$.  Suppose, for example, that all of the coefficients $\gamma_n$~, 
$C_{\ell mn}$ in Eq.~(7c) are negative; then $V_1+V_2$ is clearly 
minimized if the expectations of $\eta_n$ are all relatively real to 
one another and to the expectations of $\phi_n$, that is, if 
\begin{mathletters}
\label{allequations}
\begin{equation}
\eta_n=\pmatrix{ 0 \cr
                  \Lambda_n\cr} ~~~,
\label{equationa}
\end{equation}
with
\begin{equation}
\Lambda_1=\Lambda_2=\Lambda_3=\Omega_{\eta} ~~~,
\label{equationb}
\end{equation}
with $\Omega_{\eta}$ given by Eq.~(13b) with subscripts $\eta$ on 
all quantities.  Suppose, however, that the coefficients in $V_1$ and 
$V_2$ are all positive; then the sum $V_1+V_2$ will be made lower 
if we pick one of the degenerate minima of $V_{\eta}$ of the form of 
Eq.~(13c), for example
\begin{equation}
\Lambda_1=\overline \omega \Omega_{\eta}~,~~~ 
\Lambda_2=\omega \Omega_{\eta}~,~~~
\Lambda_3=\Omega_{\eta}~~~.
\label{equationc}
\end{equation}
\end{mathletters}
More generally, the necessary condition for Eq.~(18c) to be a lower 
minimum than Eq.~(18b), in the limit of small coupling of $\eta$ to $\phi$, 
is that $V_1+V_2$ be smaller at Eq.~(18c) than at Eq.~(18b).  Assuming 
exact cyclic permutation symmetry, which makes the following formulas 
independent of the value of the free index $m$, we find 
\begin{mathletters}
\label{allequations}
\begin{eqnarray}
V_1^{\rm Eq.(18b)}=&&3 \Omega_{\phi}^2 \Omega_{\eta}^2 
\sum_n(C_{1mn}+C_{2mn}+C_{3mn}+C_{4mn}+C_{5mn}+C_{6mn}) ~~~,\nonumber\\
V_2^{\rm Eq.(18b)}=&&3\gamma_m \Omega_{\phi} \Omega_{\eta}
+3\sum_n[(C_{7mn}+C_{8mn})\Omega_{\phi}^3\Omega_{\eta}+
(C_{9mn}+C_{10mn})\Omega_{\eta}^3\Omega_{\phi} \nonumber\\
+&&(C_{11mn}+C_{12mn})\Omega_{\phi}^2\Omega_{\eta}^2]~~~,
\label{equationa}
\end{eqnarray}
and 
\begin{eqnarray}
&&V_1^{\rm Eq.(18c)}=3\Omega_{\phi}^2 \Omega_{\eta}^2 C~~~,\nonumber\\
&&C\equiv\left(\sum_n[C_{1mn}-{1\over 2}(C_{3mn}+C_{4mn})]\right)
+C_{2mm}+C_{5m\,m+1}+C_{6m\,m+1} \nonumber\\
&&~~~-{1\over 2}[ C_{2m\,m+1} +  C_{2m\,m-1}
+C_{5mm}+C_{5m\,m-1}+C_{6mm}+C_{6m\,m-1}],\nonumber\\
&&V_2^{\rm Eq.(18c)}=0~~~.
\label{equationb}
\end{eqnarray}
Thus, the necessary condition for Eq.~(18c) to be the minimum is that 
\begin{equation}
V_1^{\rm Eq.(18c)} < V_1^{\rm Eq.(18b)}+V_2^{\rm Eq.(18b)}~~~.
\label{equationc}
\end{equation}
\end{mathletters}
We shall henceforth assume that Eq.~(19c) is satisfied; as already noted, 
this is automatic in the case when all of the coefficients in $V_1$ and 
$V_2$ are positive, but the general condition is much less restrictive, 
requiring only that the coefficients lie on one side of a hyperplane in 
the space of $V_{1,2}$ coefficients.  When Eq.~(19c) is satisfied, CP 
invariance is spontaneously broken through the $\eta$ Higgs expectations, 
and we shall see in the next section that simultaneously, the $\eta$ 
expectations have the correct form to generate nonzero second family masses.  

Let us next consider what happens when $V_1$ and $V_2$ are 
not infinitesimally 
small.  Still maintaining cyclic permutation invariance, let us first 
consider the case in which $V_1$ is large, but $V_2$ remains nearly zero. 
Then from the formulas of Appendix A, we find that the derivatives of the 
potential vanish when one assumes Eq.~(13a) for the $\phi$ expectations 
(with $\Omega$ of course replaced by $\Omega_{\phi}$) and Eq.~(18c) for 
the $\eta$ expectations, for suitable minimizing values of $\Omega_{\phi}$ 
and $\Omega_{\eta}$.  In other words, we find the correct minimum by 
first substituting  Eqs.~(13a) and (18c) into the Higgs 
potential, and then minimizing the resulting simplified expression with 
respect to $\Omega_{\phi}$ and $\Omega_{\eta}$.   Substituting 
Eqs.~(13a) and (18c) into Eq.~(7a) gives 
\begin{mathletters}
\label{allequations}
\begin{equation}
{1\over 3}{\cal L}_{\rm Higgs~potential}=A_{\phi}\Omega_{\phi}^4
-2B_{\phi}\Omega_{\phi}^2+A_{\eta}\Omega_{\eta}^4-2B_{\eta}\Omega_{\eta}^2
+C\Omega_{\phi}^2\Omega_{\eta}^2+{\rm constant}~~~,
\label{equationa}
\end{equation}
with $C$ given in Eq.~(19b), and with the remaining coefficients given by
\begin{equation}
A_{\xi}=\lambda_{\xi}-\mu_{1\xi}-\mu_{2\xi}-\alpha_{\xi},~~~
B_{\xi}=\lambda_{\xi}v_{\xi}^2,~~~\xi=\phi,\eta~~~.
\label{equationb}
\end{equation}
Minimizing Eq.~(20a) with respect to $\Omega_{\phi}^2,\Omega_{\eta}^2$ 
gives a pair of simultaneous linear equations, with the solution 
\begin{equation}
\Omega_{\phi}^2={ A_{\eta}B_{\phi}-{1 \over 2}C B_{\eta}\over
A_{\phi}A_{\eta}-{1\over 4}C^2 },~~~  
\Omega_{\eta}^2={ A_{\phi}B_{\eta}-{1 \over 2}C B_{\phi}\over
A_{\phi}A_{\eta}-{1\over 4}C^2 }~~~.
\label{equationc}
\end{equation}
In order for both $\phi$ and $\eta$ to develop nonzero vacuum expectation 
values we must have $\Omega_{\phi}^2>0,~\Omega_{\eta}^2>0$, which 
in the case when the denominator in Eq.~(20c) is positive 
requires that $C$ be restricted by 
\begin{equation}
-2A_{\phi}^{1 \over 2} A_{\eta}^{1 \over 2}< C<2 
{\rm Min}({A_{\phi} \over B_{\phi}}B_{\eta},{A_{\eta} \over B_{\eta}}
B_{\phi},A_{\phi}^{1 \over 2} A_{\eta}^{1 \over 2} )~~~. 
\label{equationd}
\end{equation}
\end{mathletters}

Because $V_1$ is invariant under independent overall phase rotations of 
$\phi$ and $\eta$, in the limit when $V_2$ is strictly zero the minimum of  
Eqs.~(13a) and (18c) is part of a one parameter $U(1)$ family of equivalent 
minima, of the form
\begin{eqnarray}
(\Omega_1,\Omega_2,\Omega_3)=&&(1,1,1) \Omega_{\phi} ~~~,\nonumber\\
(\Lambda_1,\Lambda_2,\Lambda_3)=&&(\overline \omega,\omega,1) \Omega_{\eta}
\exp(i \theta)   ~~~,
\end{eqnarray}
with the angle $\theta$ arbitrary.  When $V_2$ is nonzero but very small, the 
$U(1)$ degeneracy with respect to $\theta$ is broken, and the minimum has
the form of Eq.~(21) with a definite value of $\theta$ determined by 
the Higgs Lagrangian parameters.  A perturbative analysis in powers of 
$V_2$ shows that to first order in $V_2$ the degeneracy in $\theta$ 
is unbroken [because the final line of Eq.~(19b) remains valid for general 
$\theta$], but that at second order in $V_2$ a nontrivial condition
on $\theta$ is obtained and the degeneracy is broken.  Numerical minimization  
of the Higgs potential, using the method of Appendix A, shows that 
general values of $\theta$ can be attained at the minimum for generic   
Lagrangian parameters.  As $V_2$ 
increases, there are  relative phase and small magnitude 
corrections to the minima of Eq.~(21); when the assumption of cyclic 
permutation symmetry is relaxed, these magnitude corrections become 
more pronounced.  

To conclude our discussion of the six Higgs model, let us discuss the 
Higgs mass spectrum, assuming both exact cyclic permutation symmetry and 
the weak coupling limit in which both $V_1$ and $V_2$ are very small.    
We parameterize the expansion of $\phi_n$ and $\eta_n$ around the 
minimum as 
\begin{mathletters}
\label{allequations}
\begin{equation}
\phi_n=\pmatrix{{1\over \surd 2} \delta_n^{\phi} \cr
                  \Omega_{\phi}+{1 \over \surd 2}\epsilon_n^{\phi}\cr} ~~~,
\label{equationa}
\end{equation}                
\begin{equation}
\eta_n=\pmatrix{{1\over \surd 2} \delta_n^{\eta} \cr
                  \Omega_{\eta}+{1 \over \surd 2}\epsilon_n^{\eta}\cr} 
                  \exp(i \theta) (\overline \omega,\omega,1)_n
                  ~~~,
\label{equationb}
\end{equation}
\end{mathletters}                 
where we have used the notation $(x,y,z)_n$ to indicate $x$ for $n=1$,
$y$ for $n=2$, and $z$ for $n=3$.  Because the overall phase $\theta$ and 
the discrete chiral phases $(\overline \omega,\omega,1)_n$ drop out 
of $V_{\eta}$, for the non-Goldstone modes we get simply two copies of  
the nonzero mass modes found in Eqs.~(17a-d) in the three Higgs case, 
apart from adding subscripts or superscripts $\phi,\eta$ to distinguish 
the $\phi$ and $\eta$ sectors, as summarized in Table II.  
In computing the Yukawa couplings of the 
$\eta_n$ Higgs modes, the phases in Eq.~(22b) play a role.  Making 
transformations analogous to Eqs.~(16a) in the three Higgs case, 
with $\xi$ in the following formulas either $\phi$ or $\eta$, we have
\begin{eqnarray}
\pmatrix{\xi_1 \cr \xi_2 \cr \xi_3 \cr}=&&
W \pmatrix{\xi^{(1)} \cr\xi^{(2)} \cr \xi^{(3)} \cr}~,~~~
\pmatrix{\xi^{(1)} \cr\xi^{(2)} \cr \xi^{(3)} \cr}=
W^{-1}\pmatrix{\xi_1 \cr \xi_2 \cr \xi_3 \cr}~~~,\nonumber\\
&&\nonumber\\
\pmatrix{\delta_1^{\xi} \cr \delta_2^{\xi} \cr \delta_3^{\xi} \cr}=&&
W \pmatrix{\delta_{\xi}^{(1)} \cr\delta_{\xi}^{(2)} 
\cr \delta_{\xi}^{(3)} \cr}~,~~~
\pmatrix{\delta_{\xi}^{(1)} \cr\delta_{\xi}^{(2)} 
\cr \delta_{\xi}^{(3)} \cr}=
W^{-1}\pmatrix{\delta_1^{\xi} \cr \delta_2^{\xi} \cr \delta_3^{\xi} \cr}
~~~,\nonumber\\
&&\nonumber\\
\pmatrix{\epsilon_1^{\xi} \cr \epsilon_2^{\xi} \cr \epsilon_3^{\xi} \cr}=&&
W \pmatrix{\epsilon_{\xi}^{(1)} \cr\epsilon_{\xi}^{(2)} 
\cr \epsilon_{\xi}^{(3)} \cr}~,~~~
\pmatrix{\epsilon_{\xi}^{(1)} \cr\epsilon_{\xi}^{(2)} 
\cr \epsilon_{\xi}^{(3)} \cr}=
W^{-1}\pmatrix{\epsilon_1^{\xi} \cr \epsilon_2^{\xi} 
\cr \epsilon_3^{\xi} \cr}~~~.
\end{eqnarray}
In terms of the new bases, Eq.~(22a) becomes
\begin{mathletters}
\label{allequations}
\begin{equation}
\phi^{(n)}=\pmatrix{ {1\over \surd 2} \delta_{\phi}^{(n)}\cr 
                  {1 \over \surd 2}\epsilon_{\phi}^{(n)}\cr},~~~n=1,2,
\label{equationa}
\end{equation}
and 
\begin{equation}
\phi^{(3)}=\pmatrix{{1\over \surd 2} \delta_{\phi}^{(3)}\cr 
    \surd 3 \Omega_{\phi}+{1 \over \surd 2}\epsilon^{(3)}_{\phi}\cr}~~~,
\label{equationb}
\end{equation}
while taking into account the extra phases, Eq.~(22b) becomes
\begin{equation}
\eta^{(3)}=\pmatrix{ {1\over \surd 2} \delta_{\eta}^{(1)}\cr 
                  {1 \over \surd 2}\epsilon_{\eta}^{(1)}\cr}
                  \exp(i \theta)~~~,
\end{equation}
\begin{equation}
\eta^{(1)}=\pmatrix{ {1\over \surd 2} \delta_{\eta}^{(2)}\cr 
                  {1 \over \surd 2}\epsilon_{\eta}^{(2)}\cr}
                  \exp(i \theta)~~~,
\label{equationc}                 
\end{equation}
and 
\begin{equation}
\eta^{(2)}=\pmatrix{{1\over \surd 2} \delta_{\eta}^{(3)}\cr 
    \surd 3 \Omega_{\eta}+{1 \over \surd 2}\epsilon_{\eta}^{(3)}\cr}
                  \exp(i \theta) ~~~.
\label{equationd}
\end{equation}
\end{mathletters}
The fact that $\Omega_{\eta}$ appears in $\eta^{(2)}$ rather than in 
$\eta^{(3)}$ is directly related, as we shall see in the next section,  
to the role of the $\eta$ Higgs bosons in giving rise to second family 
masses. 

For the Goldstone modes, the situation is more complicated, because the   
$\phi$ and $\eta$ sectors interact even in the weak coupling limit.  
If $V_2$ were exactly zero, as noted above we would have 
an extra $U(1)$ symmetry, and we would get two copies of the Goldstone 
modes as well.  But for nonzero $V_2$ this $U(1)$ degeneracy is broken, and 
we are left with just one set of Goldstone modes, corresponding to the 
remaining invariance of the Higgs potential under simultaneous overall 
rephasing of $\phi,\eta$, while the three Goldstone modes related to the 
relative phase $\theta$ of $\phi$ and $\eta$ become massive pseudo-Goldstone 
modes, with squared masses that are proportional to the magnitude of $V_2$.
The decomposition of $\delta_{\phi,\eta}^{(3)}$ and 
$\epsilon_{I\phi,\eta}^{(3)}$ 
into Goldstone and pseudo-Goldstone modes is made unique by the facts   
that (i) these represent orthogonal degrees of freedom, that are simply 
rotations from the original modes 
$\delta_{\phi,\eta}^{(3)}$ and $\epsilon_{I\phi,\eta}^{(3)}$, and (ii) the 
Goldstone modes correspond precisely to a uniform 
infinitesimal phase rotation 
of $\phi,\eta$, which specifies the infinitesimal modes to which the 
pseudo-Goldstone modes must be orthogonalized.  
Since the expectations of $\phi,\eta$ may have unequal 
magnitudes $\Omega_{\phi},\Omega_{\eta}$,  we see from Eqs.~(24a-d) that 
an overall infinitesimal   
phase rotation makes a contribution to $\delta_{\eta}^{(3)}$ that is 
$\Omega_{\eta}/\Omega_{\phi}$ times as large as the corresponding 
contribution to $\delta_{\phi}^{(3)}$, and similarly makes a contribution 
to $\epsilon_{I\eta}^{(3)}$ that is $\Omega_{\eta}/\Omega_{\phi}$ times 
as large as the corresponding contribution to $\epsilon_{I\phi}^{(3)}$.
We thus find, denoting the Goldstone and   
pseudo-Goldstone modes respectively by the subscripts G and PG, and as 
before using the subscript I to denote the imaginary part,  
\begin{mathletters}
\label{allequations}
\begin{eqnarray}
\delta_G^{(3)}=&&{ \Omega_{\phi} \delta^{(3)}_{\phi} 
+ \Omega_{\eta} \delta^{(3)}_{\eta} \over 
(\Omega_{\phi}^2 + \Omega_{\eta}^2)^{1\over 2} },~~~
\delta_{PG}^{(3)}={ \Omega_{\eta} \delta^{(3)}_{\phi} 
- \Omega_{\phi} \delta^{(3)}_{\eta} \over 
(\Omega_{\phi}^2 + \Omega_{\eta}^2)^{1\over 2} }~~~,\nonumber\\
\epsilon_G^{(3)}=&&{ \Omega_{\phi} \epsilon^{(3)}_{I\phi} 
+ \Omega_{\eta} \epsilon^{(3)}_{I\eta} \over 
(\Omega_{\phi}^2 + \Omega_{\eta}^2)^{1\over 2} },~~~
\epsilon_{PG}^{(3)}={ \Omega_{\eta} \epsilon^{(3)}_{I\phi} 
- \Omega_{\phi} \epsilon^{(3)}_{I\eta} \over 
(\Omega_{\phi}^2 + \Omega_{\eta}^2)^{1\over 2} }~~~.
\label{equationa}
\end{eqnarray}
The corresponding quadratic terms in the Lagrangian are 
\begin{equation}
-{1 \over 2}[
|\partial_{\mu}\delta_G^{(3)}|^2+|\partial_{\mu}\delta_{PG}^{(3)}|^2    
+(\partial_{\mu}\epsilon_G^{(3)})^2+(\partial_{\mu}\epsilon_{PG}^{(3)})^2] 
+{1 \over 2}[M^2_{\rm charged~PG} |\delta_{PG}^{(3)}|^2    
+M^2_{\rm neutral~PG} (\epsilon_{PG}^{(3)})^2]~~~.
\label{equationb}
\end{equation}
The perturbative contribution to the pseudo-Goldstone masses, relative to 
the Higgs masses calculated above, will 
have the general magnitude (suppressing all subscripts) 
\begin{equation}
{M_{\rm PG} \over M_{\rm Higgs}} \sim 
\left( {|V_2| \over \lambda \Omega^4}\right)^{1\over 2}~~~.
\label{equationc}
\end{equation}
\end{mathletters}
We have not attempted to calculate explicit perturbative formulas for 
the pseudo-Goldstone masses, both because these will be rather complicated 
given the complexity of $V_2$ 
and because,  as argued by Weinberg [11],  there are 
likely to be significant nonperturbative corrections of order $gM_W$,  
with $g$ the electroweak gauge coupling and $M_W$ the electroweak boson 
mass.    
 
\bigskip
\section*{VI.~~~Higgs Couplings and Mass and CKM Matrices
When Cyclic Permutation Symmetry is Exact}

We proceed now to study the Yukawa couplings of the Higgs fields, and 
the mass matrices generated by their vacuum expectation values, when cyclic 
permutation symmetry is exact.  Thus, in this section we shall assume  
that the Higgs potentials have the cyclically symmetric form analyzed in 
detail in Sec. V, and we shall take the asymmetry parameters 
$\beta^f_{\xi \ell m}$ of Eqs.~(6b, c) to vanish.  As a consequence,  
the $3 \times 3$ matrices $P_{\xi n}^f$ of Eq.~(6b) are all 
retrocirculants, and are independent of the labels $\xi,f$, 
\begin{eqnarray}
P^f_{\xi 1}=&&{\rm Circ}_{\leftarrow}(0,1,0)~~~,\nonumber\\
P^f_{\xi 2}=&&{\rm Circ}_{\leftarrow}(0,0,1)~~~,\nonumber\\
P^f_{\xi 3}=&&{\rm Circ}_{\leftarrow}(1,0,0)~~~.
\end{eqnarray}
Substituting Eq.~(23) for $\xi_{1,2,3}$, with $\xi=\phi,\eta$,  
into the first line of Eq.~(6a), we get for $f=d,e$,
\begin{mathletters}
\label{allequations}
\begin{equation}
\Phi^f=
g_{\phi}^f(P_{\phi}^{f(1)}\phi^{(1)}+P_{\phi}^{f(2)}\phi^{(2)}
+P_{\phi}^{f(3)}\phi^{(3)})
+g_{\eta}^f(P_{\eta}^{f(1)}\eta^{(1)}+P_{\eta}^{f(2)}\eta^{(2)}
+P_{\eta}^{f(3)}\eta^{(3)})~~~.
\label{equationa}
\end{equation}
Here we have defined 
\begin{equation}
\pmatrix{P_{\xi}^{f(1)}\cr P_{\xi}^{f(2)}\cr P_{\xi}^{f(3)}\cr }
= W \pmatrix{P_{\xi 1}^f\cr  P_{\xi 2}^f\cr P_{\xi 3}^f \cr } 
=W^{-1} \pmatrix{P_{\xi 2}^f \cr P_{\xi 1}^f \cr P_{\xi 3}^f \cr }
~~~,
\label{equationb}
\end{equation}
\end{mathletters}
with $W$ and $W^{-1}$ as given in Eq.~(16b).  
Defining CP conjugates of $\xi^{(1,2,3)}$ by 
\begin{mathletters}
\label{allequations}
\begin{equation}
\tilde \xi^{(1,2,3)}=i \tau_2 \xi^{(1,2,3)*}~~~,
\label{equationa}
\end{equation}
the CP conjugate of the first group of equations in Eq.~(23) 
is
\begin{equation}
\pmatrix{\tilde \xi_1\cr\tilde \xi_2\cr\tilde \xi_3\cr}
=W^{-1} 
\pmatrix{\tilde \xi^{(1)}\cr\tilde \xi^{(2)}\cr\tilde\xi^{(3)} \cr  }~,~~~
\pmatrix{\tilde \xi^{(1)} \cr \tilde \xi^{(2)}\cr\tilde\xi^{(3)} \cr  }
=W \pmatrix{\tilde \xi_1\cr\tilde \xi_2\cr \tilde \xi_3 \cr }~~~.
\label{equationb}
\end{equation}
Using this for $\phi_{1,2,3},\,\eta_{1,2,3}$ in the second  
line of Eq.~(6a), we get for $f=u,\nu$, 
\begin{equation}
\Phi^f=
g_{\phi}^f(P_{\phi}^{f(1)}\tilde \phi^{(1)}+P_{\phi}^{f(2)}\tilde \phi^{(2)}
+P_{\phi}^{f(3)}\tilde \phi^{(3)})
+g_{\eta}^f(P_{\eta}^{f(1)}\tilde \eta^{(1)}+P_{\eta}^{f(2)}\tilde \eta^{(2)}
+P_{\eta}^{f(3)}\tilde \eta^{(3)})~~~.
\label{equationc}
\end{equation}
\end{mathletters}

Substituting the retrocirculant forms of Eq.~(26) into Eq.~(27b), we 
can write the matrices $P_{\xi}^{f(1,2,3)}$ as retrocirculants, 
\begin{mathletters}
\label{allequations}
\begin{eqnarray}
P_{\xi}^{f(1)}=&&{1 \over \surd 3}
{\rm Circ}_{\leftarrow}(1,\omega,\overline \omega)~~~,\nonumber\\
P_{\xi}^{f(2)}=&&{1 \over \surd 3} 
{\rm Circ}_{\leftarrow}(1,\overline\omega,\omega)~~~,\nonumber\\
P_{\xi}^{f(3)}=&&{1 \over \surd 3}
{\rm Circ}_{\leftarrow}(1,1,1)~~~.
\label{equationa}
\end{eqnarray}
Let us now use Eq.~(10b), which asserts that $P_{\xi}^{f(1,2,3)}$ are all 
diagonalized by the same bi-unitary transformation constructed 
using $U_L,U_R^{\dagger}$ of Eq.~(10a), 
\begin{eqnarray}
U_LP_{\xi}^{f(1)}U_R^{\dagger}=&&\surd 3  {\rm diag} (1,0,0) \equiv 
\surd 3M^{(1)}~~~,\nonumber\\
U_LP_{\xi}^{f(2)}U_R^{\dagger}=&&\surd 3 {\rm diag} (0,1,0) \equiv
\surd 3M^{(2)} ~~~,\nonumber\\
U_LP_{\xi}^{f(3)}U_R^{\dagger}=&&\surd 3 {\rm diag} (0,0,1)\equiv
\surd 3 M^{(3)} ~~~.
\label{equationb}
\end{eqnarray}
\end{mathletters}
Clearly, the natural thing to do now is to rotate to new fermion bases 
using the same matrices $U_L,U_R$, by introducing primed bases defined by 
\begin{mathletters}
\label{allequations}
\begin{eqnarray}
Q_L=&&U_L^{\dagger} Q_L^{\prime},~\psi_L=U_L^{\dagger}
\psi_L^{\prime}~~~,\nonumber\\ 
f_R=&&U_R^{\dagger}f_R^{\prime},~~f=d,u,e,\nu~~~.
\label{equationa}
\end{eqnarray}
Since the fermion kinetic energy of Eq.~(4c) does not couple left to right  
chiral components, it has the same form in terms of 
the primed bases as in terms of the original ones.  Substituting 
Eqs.~(27a), (28c), and (29b) into the Yukawa Lagrangian of Eq.~(5b), 
we get finally 
\begin{equation}
{\cal L}_{\rm Yukawa}=\overline Q_L^{\,\prime} \Psi^d d_R^{\prime} 
+\overline Q_L^{\,\prime} \Psi^u u_R^{\prime} + \overline \psi_L^{\,\prime} 
\Psi^e e_R^{\prime} +\overline \psi_L^{\,\prime} \Psi^{\nu} \nu_R^{\prime}
~~~+{\rm adjoint}~~~,
\label{equationb}
\end{equation}
with the $3 \times 3$ matrices $\Psi^f$ defined by 
\begin{eqnarray}
\Psi^f=&&\sum_{\ell=1}^3 \surd 3
(g_{\phi}^f\phi^{(\ell)}+g_{\eta}^f\eta^{(\ell)})M^{(\ell)},
~~~f=d,e~~~,\nonumber\\
\Psi^f=&&\sum_{\ell=1}^3 \surd 3
(g_{\phi}^f\tilde \phi^{(\ell)}+g_{\eta}^f\tilde \eta^{(\ell)})
M^{(\ell)},
~~~f=u,\nu~~~.
\label{equationc}
\end{eqnarray}
\end{mathletters}

On substituting Eqs.~(24a-d) into Eq.~(30c), we can read off both the 
mass matrices and the Yukawa couplings of the physical Higgs states.  
The mass matrices are obtained by keeping only the vacuum expectations 
of $\phi^{(\ell)},\eta^{(\ell)}$, that is, by setting 
\begin{mathletters}
\label{allequations}
\begin{equation}
\phi^{(1,2)}\to 0,~ \tilde \phi^{(1,2)} \to 0,~
\phi^{(3)} \to \pmatrix{ 0 \cr \surd 3 \Omega_{\phi} \cr},~
\tilde \phi^{(3)} \to \pmatrix{ \surd 3 \Omega_{\phi} \cr 0 \cr}~~~~,
\label{equationa}
\end{equation}
and 
\begin{equation}
\eta^{(1,3)}\to 0,~ \tilde \eta^{(1,3)} \to 0,~
\eta^{(2)} \to \pmatrix{ 0 \cr \surd 3 \Omega_{\eta} \exp(i\theta)\cr},~
\tilde \eta^{(2)} \to \pmatrix{ \surd 3 \Omega_{\eta} \exp(-i\theta)
\cr 0 \cr}~~~~,
\label{equationb}
\end{equation}
\end{mathletters}
giving
\begin{mathletters}
\label{allequations}
\begin{eqnarray}
{\cal L}_{\rm mass}=&&
\overline d^{\,\prime}_L(3g_{\eta}^d\Omega_{\eta}\exp(i\theta)M^{(2)}
+3g_{\phi}^d\Omega_{\phi}M^{(3)})d_R^{\prime} \nonumber\\
+&&\overline u^{\,\prime}_L(3g_{\eta}^u\Omega_{\eta}\exp(-i\theta)M^{(2)}
+3g_{\phi}^u\Omega_{\phi}M^{(3)})u_R^{\prime} \nonumber\\
+&&\overline e^{\,\prime}_L(3g_{\eta}^e\Omega_{\eta}\exp(i\theta)M^{(2)}
+3g_{\phi}^e\Omega_{\phi}M^{(3)})e_R^{\prime} \nonumber\\
+&&\overline {\nu}^{\,\prime}_L(3g_{\eta}^{\nu}\Omega_{\eta}
\exp(-i\theta)M^{(2)}
+3g_{\phi}^{\nu}\Omega_{\phi}M^{(3)})\nu_R^{\prime}
~~~+{\rm adjoint}~~~. 
\label{equationa}
\end{eqnarray}
Identifying  $M^{(1,2,3)}$  respectively as the projectors on the 
first, second, and third family states in the primed basis, we 
read off from Eq.~(32a) the masses 
\begin{eqnarray}
M_t=&&3g_{\phi}^u
\Omega_{\phi},~M_c=3g_{\eta}^u\Omega_{\eta},~M_u=0~~~,\nonumber\\ 
M_b=&&3g_{\phi}^d
\Omega_{\phi},~M_s=3g_{\eta}^d\Omega_{\eta},~M_d=0~~~,\nonumber\\ 
M_{\tau}=&&3g_{\phi}^e \Omega_{\phi},~M_{\mu}=3g_{\eta}^e
\Omega_{\eta},~M_e=0~~~,\nonumber\\
M_{\nu_{\tau}}=&&3g_{\phi}^{\nu} \Omega_{\phi},~
M_{\nu_{\mu}}=3g_{\eta}^{\nu}\Omega_{\eta},~M_{\nu_e}=0~~~.
\label{equationb}
\end{eqnarray}
\end{mathletters}
We see that in the three Higgs model, only the third family gets masses, 
with the first two families remaining massless.  The same is true in the 
CP conserving phase of the six Higgs model, in which the $\eta$ expectations 
are given by Eq.~(18b) rather than Eq.~(18c); in this phase, the projectors 
$M^{(2)}$ in Eq.~(32a) are replaced by projectors $M^{(3)}$, and the 
$\eta$ expectations simply make additional contributions to the third 
family masses.  On the other hand, in the phase of the six Higgs model that 
spontaneously violates CP as in Eqs. (18c) and (21), the factors  
$\overline \omega, \omega, 1$ in Eq.~(18c) 
give rise to the projector $M^{(2)}$ for the 
second family states, which then receive masses.  The hierarchy between 
the masses of the second and third family charged leptons is attributed, 
in the six Higgs model, to a systematic tendency of the $\eta$ Higgs 
bosons to have smaller Yukawa couplings to the charged fermions than 
those of the $\phi$ Higgs bosons.  

To get a feeling for the magnitudes involved, we note that the Higgs boson 
expectations generate mass terms for the gauge bosons given by  
\begin{mathletters}
\label{allequations}
\begin{equation}
{\cal L}_{\rm gauge ~mass}=
[-{g^2 \over 4}W_{+\mu}W_{-\mu}-{1\over 8}(gW_{3\mu}-g^{\prime}B_{\mu})^2]
v^2~~~,
\label{equationa}
\end{equation}
with 
\begin{equation}
v^2=2\sum_{n=1}^3(|\langle \phi_n \rangle|^2 +|\langle \eta_n \rangle|^2)
=6(\Omega_{\phi}^2 + \Omega_{\eta}^2)~~~.
\label{equationb}
\end{equation}
\end{mathletters}
Empirically, $v \simeq 247 {\rm GeV}$; assuming, as we shall in the fits 
below, that $\Omega_{\phi}$ and $\Omega_{\eta}$ are approximately equal, we
then find $\Omega_{\phi} \simeq \Omega_{\eta}\simeq 71 {\rm GeV}$.  
The Yukawa couplings needed to reproduce the observed charged fermion 
masses are then given in the six Higgs model by 
\begin{mathletters}
\label{allequations}
\begin{eqnarray}
g_{\phi}^u\simeq && 0.81,~g_{\eta}^u\simeq 0.0061~~~,\nonumber\\
g_{\phi}^d\simeq && 0.020,~g_{\eta}^d\simeq 0.00094~~,\nonumber\\
g_{\phi}^e\simeq && 0.0083,~g_{\eta}^e\simeq 0.00050~~~.
\label{equationa}
\end{eqnarray}
In the three Higgs model, $\Omega_{\phi}$ is a factor $\surd 2$ larger 
than in the six Higgs model, and the $\phi$ Yukawa couplings are 
correspondingly a factor $\surd 2$ smaller than in Eq.~(34a),  
\begin{equation}
g_{\phi}^u\simeq 0.57,~g_{\phi}^d\simeq 0.014,~g_{\phi}^e \simeq 0.0059
\label{equationb}
\end{equation}
\end{mathletters}

As we have seen, because the mass matrices in the cyclically symmetric 
limit are retrocirculants, we were able to diagonalize them with 
universal, flavor independent matrices $U_L,\,U_R$.  This has the 
important consequence that when cyclic symmetry is assumed as a 
leading approximation, the corresponding approximation to the 
CKM mixing matrix is unity, a welcome feature since the observed  
CKM matrix is close to unity.   A related welcome feature of the cyclic
approximation is that there are no flavor changing neutral currents,  
which again accords with the fact that these are observed to be highly    
suppressed.  To obtain realistic non-unit  
values for the CKM matrix, we will have to go beyond the cyclic approximation 
by including nonzero asymmetries $\beta_{\xi \ell m}^f$ as in Eq.~(6b), 
but we shall then also have to estimate the magnitude of the flavor changing  
neutral current effects produced by these asymmetries.  This will be the 
agenda of the next three sections.  

Before proceeding with this analysis, we note that the leading cyclic 
approximation to the Yukawa couplings of the physical Higgs bosons can 
be read off from Eqs.~(30b, c) together with Eqs.~(24a-d) and (25a).  
We see that 
\begin{mathletters}
\label{allequations}
\begin{eqnarray}
&&\delta_{\phi}^{(1)},~\epsilon_{\phi}^{(1)},~\delta_{\eta}^{(2)},~
\epsilon_{\eta}^{(2)} ~~{\rm couple~only~to~the~first~family}~~~,\nonumber\\
&&\delta_{\phi}^{(2)},~\epsilon_{\phi}^{(2)},~\delta_{\eta}^{(3)},~
\epsilon_{\eta}^{(3)} ~~{\rm couple~only~to~the~second~family}~~~,\nonumber\\
&&\delta_{\phi}^{(3)},~\epsilon_{\phi}^{(3)},~\delta_{\eta}^{(1)},~
\epsilon_{\eta}^{(1)} ~~{\rm couple~only~to~the~third~family}~~~,
\label{equationa}
\end{eqnarray}
which by Eq.~(17c) imply that 
\begin{eqnarray}
&&\epsilon_{\phi R,I}^{(\pm)} ~~{\rm couple~only~to~the~first~
and~second~families}~~~,\nonumber\\
&&\epsilon_{\eta R,I}^{(\pm)} ~~{\rm couple~only~to~the~first~
and~third~families}~~~,
\label{equationb}
\end{eqnarray}
and by Eq.~(25a) imply that 
\begin{equation}
\delta_{PG}^{(3)},~\epsilon_{PG}^{(3)} ~~{\rm couple~only~to~the~second~ 
and~third~families}~~~.
\label{equationc}
\end{equation}
\end{mathletters}
The presence of 21 Higgs bosons in the six Higgs doublet model 
[eight charged Higgs bosons $\delta_{\phi,\eta}^{(1,2)}$, ten   
neutral Higgs bosons 
$\epsilon_{R \phi,\eta}^{(3)},\,\epsilon_{R \phi,\eta}^{(\pm)}$, 
and $\epsilon_{I \phi,\eta}^{(\pm)}$, plus two charged pseudo-Goldstone Higgs 
bosons $\delta_{PG}^{(3)}$, and one neutral pseudo-Goldstone 
Higgs boson $\epsilon_{PG}^{(3)}$], together with the 
pattern of predominant fermionic 
couplings given in Eqs.~(35a-c) and summarized in Table II, 
are a distinguishing feature of the 
model that should be testable in experiments at 
the next generation of accelerators. 

\section*{ VII.~~~First Order Breaking of Cyclic Permutation Symmetry} 

We now set up a perturbative scheme to study the effects of the breaking 
of cyclic permutation symmetry.  In the three Higgs model, we will also 
allow CP noninvariance of the Lagrangian, by allowing 
the phases $\psi$ in Eq.~(7b) to be 
nonzero, and by allowing the Yukawa couplings to be complex.  In the 
six Higgs model, we will impose CP invariance on the Lagrangian, but will 
work in the phase that spontaneously breaks CP.  Two types of first order 
small corrections will be introduced. The first are corrections to the  
Higgs vacuum expectations, arising from a lack of cyclic symmetry in the 
Higgs potential.  In the three Higgs model, this results in replacing 
Eq.~(13a) by 
\begin{mathletters}
\label{allequations}
\begin{equation}
\Omega_n=\Omega(1+\delta_n),~~n=1,2,3~~~,
\label{equationa}
\end{equation}
where the $\delta_n$ are small corrections that can be complex, and where 
we impose the condition 
\begin{equation}
\sum_n \delta_n=0~~~
\label{equationb}
\end{equation}
\end{mathletters}
to avoid duplicating information contained in the overall factor $\Omega$ 
and the overall phase that has been eliminated by a gauge transformation.  
In the six Higgs model, we have analogous 
corrections to the first line in Eq.~(21),  
\begin{equation}
\Omega_n=\Omega_{\phi}(1+\delta_n),~~n=1,2,3,~\sum_n\delta_n=0~~~,    
\end{equation}
where the $\delta_n$ can again be complex when the potentials $V_1,V_2$ 
that couple $\phi$ to $\eta$ are not neglected.  In principle, there 
are also asymmetry corrections to the second line of Eq.~(21), which gives 
the $\eta$ expectations. But these are always suppressed by a factor  
$g_{\eta}^f/g_{\phi}^f$,  which according to Eq.~(34a) is at most of order
0.06, and so will be neglected in what follows; that is, we treat $g_{\eta}
/g_{\phi}$ here as if it were also a first order small quantity.  The 
second type of first order small corrections are the asymmetry parameters 
$\beta_{\phi mn}^f$ of Eqs.~(6b, c), which are complex in the three 
Higgs model when explicit CP violation is permitted, but are real in the 
six Higgs model when CP invariance is imposed on the Lagrangian.  Again, 
in principle there are analogous asymmetry parameters 
$\beta_{\eta mn}^f$ for the $\eta$ Yukawa couplings, but the effect 
of these is again suppressed by a factor $g_{\eta}^f/g_{\phi}^f$ and so they 
will be neglected.  This itemization of corrections defines the model that 
we shall study in first order perturbation theory.  

Since the zeroth order problem, that was analyzed in Sec. VI, is brought 
to diagonal form by the bi-unitary transformations of Eqs. (29b) and (30a) 
based on the matrices $U_L,U_R$ of Eq.~(10a), we shall make this 
transformation at the outset. In the primed fermion basis, the zeroth 
order mass matrix is still given by Eq.~(32a), but now there will be 
first order corrections from the $\delta$'s and $\beta$'s introduced above.   
Since we are regarding $g_{\eta}^f/g_{\phi}^f$ as effectively a first order 
correction, it is convenient to group it with the other first order terms.  
Starting again from Eqs.~(23), (27a, b), and (28c), we then 
find for the extension of Eq.~(32a) to include all first order 
corrections, 
\begin{mathletters}
\label{allequations}
\begin{equation}
{\cal L}_{\rm mass}=\sum_{f=d,u,e,\nu}
\overline f_L^{\,\prime}g_{\phi}^f\Omega_{\phi}
(3M^{(3)}+\sigma^f)f_R^{\prime}~~~,
\label{equationa}
\end{equation}
with $\sigma^f$ a $3 \times 3$ matrix with matrix elements given by 
\begin{eqnarray}
\sigma_{11}^f=&&{1 \over 3} \mu_{11}^f 
+\delta_3^f+\overline \omega \delta_2^f +\omega \delta_1^f  ~~~,\nonumber\\
\sigma_{22}^f=&&{1\over 3} \mu_{22}^f + 3 R^f 
+\delta_3^f+ \omega \delta_2^f +\overline\omega \delta_1^f  ~~~,\nonumber\\
\sigma_{33}^f=&&0~~~,\nonumber\\
\sigma_{\ell m}^f=&&{1\over 3} \mu_{\ell m}^f,~\ell \not= m~~~.
\label{equationb}
\end{eqnarray}
\end{mathletters}
The further quantities appearing in Eq.~(38b) are defined as follows.  
The quantities $\delta_n^f$ are given, 
in terms of the $\delta_n$ introduced in Eqs.~(36) and (37), by 
\begin{mathletters}
\label{allequations}
\begin{eqnarray}
\delta_1^f=&&\delta_2,~~\delta_2^f=\delta_1,~~\delta_3^f=\delta_3,
~~f=d,e~~~,\nonumber\\
\delta_1^f=&&\delta_1^*,~~\delta_2^f=\delta_2^*,~~\delta_3^f=\delta_3^*,~~
f=u,\nu~~~.
\label{equationa}
\end{eqnarray}
The quantities 
$R^f$ are defined by 
\begin{equation}
R^f={g_{\eta}^f \Omega_{\eta} \exp(\pm i \theta) \over 
g_{\phi}^f \Omega_{\phi} }~~~,
\label{equationb}
\end{equation}
with the $+$ sign holding for $f=d,e$ and the $-$ sign holding for 
$f=u,\nu$.  Finally, the $\mu_{\ell m}^f$'s, when multiplied by the factor  
of $1/3$ in Eq.~(38b), are the 
asymmetries $\beta_{\phi \ell m}^f$ reexpressed in the primed fermion basis; 
suppressing the subscript $\phi$ on the $\beta$'s, they are given by  
\begin{eqnarray}
\mu_{11}^f=&&\beta_{11}^f+\beta_{23}^f+\beta_{32}^f 
+\overline \omega (\beta_{12}^f+\beta_{21}^f+\beta_{33}^f)
+\omega (\beta_{13}^f + \beta_{22}^f + \beta_{31}^f)~~~,\nonumber\\
\mu_{22}^f=&&\beta_{11}^f+\beta_{23}^f+\beta_{32}^f 
+ \omega (\beta_{12}^f+\beta_{21}^f+\beta_{33}^f)
+\overline\omega (\beta_{13}^f + \beta_{22}^f + \beta_{31}^f)~~~,\nonumber\\
&&\phantom{\vrule height 25pt}\nonumber\\
\mu_{12}^f=&&\beta_{11}^f+\beta_{22}^f+\beta_{33}^f 
+ \omega (\beta_{12}^f+\beta_{23}^f+\beta_{31}^f)
+\overline\omega (\beta_{21}^f + \beta_{32}^f + \beta_{13}^f)~~~,\nonumber\\
\mu_{21}^f=&&\beta_{11}^f+\beta_{22}^f+\beta_{33}^f 
+ \overline \omega (\beta_{12}^f+\beta_{23}^f+\beta_{31}^f)
+ \omega (\beta_{21}^f + \beta_{32}^f + \beta_{13}^f)~~~,\nonumber\\
&&\phantom{\vrule height 25pt}\nonumber\\
\mu_{13}^f=&&\beta_{11}^f+\beta_{12}^f+\beta_{13}^f 
+\overline \omega (\beta_{21}^f+\beta_{22}^f+\beta_{23}^f)
+\omega (\beta_{31}^f + \beta_{32}^f + \beta_{33}^f)~~~,\nonumber\\
\mu_{23}^f=&&\beta_{11}^f+\beta_{12}^f+\beta_{13}^f 
+ \omega (\beta_{21}^f+\beta_{22}^f+\beta_{23}^f)
+\overline\omega (\beta_{31}^f + \beta_{32}^f +
\beta_{33}^f)~~~,\nonumber\\
&&\phantom{\vrule height 25pt}\nonumber\\
\mu_{31}^f=&&\beta_{11}^f+\beta_{21}^f+\beta_{31}^f 
+\overline \omega (\beta_{12}^f+\beta_{22}^f+\beta_{32}^f)
+\omega (\beta_{13}^f + \beta_{23}^f + \beta_{33}^f)~~~,\nonumber\\
\mu_{32}^f=&&\beta_{11}^f+\beta_{21}^f+\beta_{31}^f 
+ \omega (\beta_{12}^f+\beta_{22}^f+\beta_{32}^f)
+\overline\omega (\beta_{13}^f + \beta_{23}^f + \beta_{33}^f)~~~.
\label{equationc}
\end{eqnarray}
\end{mathletters}
We remark that since CP invariance requires the $\beta$'s to be real, 
the condition for CP invariance, when expressed directly  
in terms of the $\mu$'s, is $\mu_{11}^{f*}=\mu_{22}^f$, 
$\mu_{12}^{f*}=\mu_{21}^f$, $\mu_{13}^{f*}=\mu_{23}^f$, and
$\mu_{31}^{f*}=\mu_{32}^f$.  

Defining 
\begin{mathletters}
\label{allequations}
\begin{equation}
M_f^{\prime }\equiv 3M^{(3)}+\sigma^f ~~~,
\label{equationa}
\end{equation}
we must now find the bi-unitary transformation matrices 
$U_L^f,U_R^f$ for which $U_L^f M_f^{\prime} U_R^{f\dagger}$ is diagonal, 
with the eigenvalues ordered in absolute value, for each flavor $f$.  The  
fermion basis states that are mass eigenstates   
are then related to the primed basis by 
\begin{eqnarray}
f_L^{\prime}=&&U_L^{f\dagger} f_L^{\rm mass}~~~,\nonumber\\
f_R^{\prime}=&&U_R^{f\dagger} f_R^{\rm mass},~~f=d,u,e,\nu~~~,
\label{equationb}
\end{eqnarray}
and the CKM matrix $U_{\rm CKM}$ is given as usual by 
\begin{equation}
U_{\rm CKM}=U_L^{u\dagger}U_L^d~~~.
\label{equationc}
\end{equation}
\end{mathletters}

We shall now develop a perturbative procedure for calculating $U_{L,R}^f$.  
The first observation to be made is that we are dealing with a degenerate 
perturbation problem, since the zeroth order mass matrix $3M^{(3)}
=3{\rm diag}(0,0,1)$ has eigenvalues 0 for the first two primed basis 
states.  As a consequence,  the $2 \times 2$ submatrix of 
$U_{L,R}^f$ spanned by these states is zeroth order in the perturbation 
$\sigma^f$, with only the off-diagonal elements coupling to the third basis 
state of first order.  Thus, we find a natural reason in our model why the 
CKM mixings of the first and second family states should be larger than 
the mixings of the first and second families with the third family.  

We shall deal with the zeroth order $2 \times 2$ submatrix by calculating  
it exactly.  Let $V_{L,R}^f$ be the $2 \times 2$ matrices that bring 
the $2 \times 2$ submatrix of $\sigma^f$ to diagonal form,
\begin{mathletters}
\label{allequations}
\begin{equation}
V_L^f \pmatrix{ \sigma_{11}^f & \sigma_{12}^f \cr
\sigma_{21}^f & \sigma_{22}^f \cr} V_R^{f\dagger}
=\pmatrix{ \kappa_1^f & 0\cr 0 & \kappa_2^f\cr  }~~~,
\label{equationa}
\end{equation}
with the magnitudes of the 
eigenvalues ordered as $|\kappa_1^f| \leq |\kappa_2^f|$.
The explicit construction of $V_{L,R}^f$ is given in Appendix B.   
It is then straightforward to show that to first order in small 
quantities, $U_{L,R}^f$ are given by 
\begin{equation}
U_L^f=\pmatrix{ V_L^f & -{1\over 3}V_L^f 
\pmatrix{\sigma_{13}^f \cr\sigma_{23}^f \cr} \cr
{1\over 3} \pmatrix{\sigma_{13}^f \cr\sigma_{23}^f \cr}^{\dagger} & 1 \cr }
\label{equationb}
\end{equation}
\begin{equation}
U_R^{f\dagger}=\pmatrix{ V_R^{f\dagger} & {1\over 3} 
\pmatrix{\sigma_{31}^{f*} \cr\sigma_{32}^{f*} \cr} \cr
-{1\over 3} \pmatrix{\sigma_{31}^{f*} \cr\sigma_{32}^{f*} \cr}^{\dagger} 
 V_R^{f\dagger}     & 1 \cr }
\label{equationc}
\end{equation}
and 
\begin{equation}
U_L^f M_f^{\prime} U_R^{f\dagger}=\pmatrix{\kappa_1^f &0&0\cr
0&\kappa_2^f&0\cr 0&0&3 \cr}~~~.
\label{equationd}
\end{equation}
\end{mathletters}
Defining 
\begin{mathletters}
\label{allequations}
\begin{equation}
V_{\rm CKM} \equiv   V_L^{u\dagger}V_L^d~~~,
\label{equationa}
\end{equation}
the corresponding first order accurate expression for the CKM matrix 
is given by 
\begin{equation}
U_{\rm CKM}=\pmatrix{ V_{\rm CKM} & 
-{1\over 3} V_{\rm CKM} \pmatrix{\sigma_{13}^d \cr \sigma_{23}^d \cr } 
+ {1 \over 3} \pmatrix{\sigma_{13}^u \cr \sigma_{23}^u \cr }  \cr
{1\over 3} \pmatrix{\sigma_{13}^d \cr \sigma_{23}^d \cr }^{\dagger} 
- {1 \over 3} \pmatrix{\sigma_{13}^u \cr \sigma_{23}^u \cr }^{\dagger} 
V_{\rm CKM}  
&1\cr }~~~.
\label{equationb}
\end{equation}
\end{mathletters}
Although Eq.~(42b) is useful for analytic study of the CKM matrix, in our 
numerical work we shall simply compute directly from the definition of  
Eq.~(40c).  We shall also, in the numerical work,  
use slightly more accurate forms for $U_{L,R}^f$
in which only the square of the ratio of second to third family 
masses $(|\kappa_2^f|/3)^2$ is assumed to be small; 
the relevant formulas are given in Appendix C.   

\section*{VIII.~~~Higgs Exchange Contributions to the $K_L-K_S$ 
Mass Difference} 

As pointed out in Sec. VI, when cyclic  symmetry is exact, Higgs boson  
exchange in our models does not produce strangeness changing neutral current 
effects.  However, once we include cyclic asymmetries, such effects become 
possible and we must be sure that their magnitude does not exceed known 
experimental limits.  Since, in the context of extensions of the Higgs 
sector, the most stringent bound on strangeness changing neutral current 
processes comes [13] from the second order weak $K_L-K_S$ mass difference, 
we shall consider only this process, and shall calculate the contribution 
to its matrix element arising from Higgs exchange within the perturbative 
framework set up in Sec. VII.  

We saw there that, because the zeroth 
order mass matrix is degenerate in the subspace spanned by the first two 
families, the mixing matrices within this subspace are {\it zeroth order} 
rather than first order in the perturbation, and therefore strangeness 
changing neutral current effects can already appear at zeroth order in 
perturbation theory.  What we shall do in this section is to calculate 
this zeroth order contribution to the $K_L-K_S$ mass difference, neglecting 
all terms of first and higher order in the asymmetric perturbation.  Our 
starting point is thus the Yukawa Lagrangian of Eqs.~(30b, c) in the 
primed basis, of which the term relevant to mixing of the $d$ and $s$ quarks 
is  
\begin{mathletters}
\label{allequations}
\begin{equation}
\overline Q_L^{\,\prime}\sum_{\ell=1}^3 \surd 3 (g_{\phi}^d \phi^{(\ell)} 
+ g_{\eta}^d \eta^{(\ell)}) M^{(\ell)}d_R^{\prime}~
+~{\rm adjoint}  ~~~.
\label{equationa}
\end{equation}
Substituting Eq.~(40b) relating the primed to the mass eigenstate bases and 
using the approximation of Eqs.~(41b, c) for $U_{L,R}^d$;  also 
substituting Eqs.~(24a, b) for the $\phi^{(\ell)}$ and keeping only the 
neutral Higgs pieces; and finally also neglecting terms of first and 
higher order in the asymmetric perturbation, we get the effective Lagrangian 
\begin{equation}
{\cal L}_{\rm scnc}\equiv
\overline d_L^{\rm\, mass}{\surd 3 \over \surd 2} g_{\phi}^d 
\sum_{\ell=1}^2 \epsilon_{\phi}^{(\ell)}V_L^d M^{(\ell)}_{2 \times 2}
V_R^{d\dagger} d_R^{\rm mass}~+~{\rm adjoint}~~~.
\label{equationb}
\end{equation}
In Eq.~(43b),  the  subscript $2 \times 2$ on the projectors indicates their 
restriction to the subspace spanned by the first two families, and the   
column vector $d$ will be understood to have been truncated from three 
to two components, corresponding to the first two families.  Finally, 
reexpressing $\epsilon_{\phi}^{(1,2)}$
in terms of the modes $\epsilon_{\phi}^{(\pm)}$ 
defined in Eq.~(17c), splitting these into real and imaginary parts,  
and explicitly including the adjoint term (our 
$\gamma$ matrix conventions are $\gamma_5=\gamma_5^{\dagger}$, 
$\gamma^0=\gamma^{0\dagger}$, $(\gamma^0)^2=1$), we get
\begin{eqnarray}
{\cal L}_{\rm scnc} 
 =&&\overline d^{\rm\, mass} {\surd 3 \over 4} g_{\phi}^d \{
[(\epsilon_{\phi R}^{(+)}+\epsilon_{\phi R}^{(-)})+i
(\epsilon_{\phi I}^{(+)}+\epsilon_{\phi I}^{(-)})]V_L^d 
M^{(1)}_{2\times 2} V_R^{d\dagger}\nonumber\\
+&&[(\epsilon_{\phi R}^{(+)}-\epsilon_{\phi R}^{(-)})+i
(\epsilon_{\phi I}^{(+)}-\epsilon_{\phi I}^{(-)})]V_L^d M^{(2)}_{2\times 2} 
V_R^{d\dagger} \} (1+\gamma_5) d^{\rm mass} \nonumber\\
 +&&\overline d^{\rm\, mass} {\surd 3 \over 4} g_{\phi}^{d*} \{
[(\epsilon_{\phi R}^{(+)}+\epsilon_{\phi R}^{(-)})-i
(\epsilon_{\phi I}^{(+)}+\epsilon_{\phi I}^{(-)})]V_R^d 
M^{(1)}_{2\times 2} V_L^{d\dagger} \nonumber\\
+&&[(\epsilon_{\phi R}^{(+)}-\epsilon_{\phi R}^{(-)})-i
(\epsilon_{\phi I}^{(+)}-\epsilon_{\phi I}^{(-)})]V_R^d 
M^{(2)}_{2\times 2} V_L^{d\dagger} \} (1-\gamma_5) d^{\rm mass} ~~~.
\label{equationc}
\end{eqnarray}
\end{mathletters}
To facilitate the remaining calculation, it is convenient to rewrite 
Eq.~(43c) in the form 
\begin{mathletters}
\label{allequations}
\begin{eqnarray}
{\cal L}_{\rm scnc}
=&&\overline d^{\rm\, mass} \epsilon_{\phi R}^{(+)}
(A_R^{(+)}+B_R^{(+)}\gamma_5) d^{\rm mass} 
+ \overline d^{\rm\, mass} \epsilon_{\phi R}^{(-)}
(A_R^{(-)}+B_R^{(-)}\gamma_5) d^{\rm mass} \nonumber\\
+&&\overline d^{\rm\, mass} \epsilon_{\phi I}^{(+)}
(A_I^{(+)}+B_I^{(+)}\gamma_5) d^{\rm mass} 
+ \overline d^{\rm\, mass} \epsilon_{\phi I}^{(-)}
(A_I^{(-)}+B_I^{(-)}\gamma_5) d^{\rm mass}~~~. 
\label{equationa}
\end{eqnarray}
Using the facts that
\begin{equation}
M^{(1)}_{2\times 2}+M^{(2)}_{2\times 2}=\pmatrix{1&0\cr0&1\cr}\equiv 1,~~ 
  M^{(1)}_{2\times 2}-M^{(2)}_{2\times 2}=\pmatrix{1&0\cr0&-1\cr}
  \equiv \rho_3~~~, 
\label{equationb}
\end{equation}
we find that the $2\times 2$ matrices $A_{R,I}^{(\pm)}$, $B_{R,I}^{(\pm)}$ 
appearing in Eq.~(44a) are given by 
\begin{eqnarray}
A_R^{(+)}=&&{\surd 3 \over 4} g_{\phi}^d V_L^d V_R^{d\dagger}       
+         {\surd 3 \over 4} g_{\phi}^{d*} V_R^d V_L^{d\dagger},~~       
B_R^{(+)}={\surd 3 \over 4} g_{\phi}^d V_L^d V_R^{d\dagger}       
-         {\surd 3 \over 4} g_{\phi}^{d*} V_R^d V_L^{d\dagger},~~ \nonumber\\
A_R^{(-)}=&&{\surd 3 \over 4} g_{\phi}^d V_L^d \rho_3 V_R^{d\dagger}       
+         {\surd 3 \over 4} g_{\phi}^{d*} V_R^d \rho_3 V_L^{d\dagger},~~       
B_R^{(-)}={\surd 3 \over 4} g_{\phi}^d V_L^d \rho_3 V_R^{d\dagger}       
-         {\surd 3 \over 4} g_{\phi}^{d*} V_R^d \rho_3
V_L^{d\dagger},~~ \nonumber\\       
&&\phantom{\vrule height 20pt}\\
A_I^{(+)}=&&{\surd 3 \over 4} g_{\phi}^d i V_L^d V_R^{d\dagger}       
-         {\surd 3 \over 4} g_{\phi}^{d*} i V_R^d V_L^{d\dagger},~~       
B_I^{(+)}={\surd 3 \over 4} g_{\phi}^d i V_L^d V_R^{d\dagger}       
+         {\surd 3 \over 4} g_{\phi}^{d*} i V_R^d V_L^{d\dagger},~~\nonumber\\
A_I^{(-)}=&&{\surd 3 \over 4} g_{\phi}^d i V_L^d \rho_3 V_R^{d\dagger}       
-         {\surd 3 \over 4} g_{\phi}^{d*} i V_R^d \rho_3 V_L^{d\dagger},~~
B_I^{(-)}={\surd 3 \over 4} g_{\phi}^d i V_L^d \rho_3 V_R^{d\dagger}       
+         {\surd 3 \over 4} g_{\phi}^{d*} i V_R^d \rho_3 V_L^{d\dagger}~~~. 
\label{equationc}
\end{eqnarray}
\end{mathletters}
Letting $d$ and $s$ denote, respectively, 
the down and strange quark eigenstates, the two component column vector 
$d^{\rm mass}$ has the structure 
\begin{mathletters}
\label{allequations}
\begin{equation}
d^{\rm mass}=\pmatrix{d \cr s}~~~,
\label{equationa}
\end{equation}
and so for any $2 \times 2$ matrix $N$, we have 
\begin{equation}
\overline d^{\rm \,mass} N d^{\rm mass}=\overline d N_{11} d 
+\overline d N_{12} s + \overline s N_{21} d + \overline s N_{22} s~~~.
\label{equationb}
\end{equation}
Hence the strangeness changing terms of Eq.~(44a) involve only the 12 and 
21 matrix elements of the matrices in Eq.~(44c), and can be compactly 
written as 
\begin{equation}
{\cal L}_{\rm scnc}^{\Delta S=1}
=\sum_{p=\pm}\sum_{F=R,I}
[\overline d \epsilon_{\phi F}^{(p)} (A_{F12}^{(p)}+B_{F12}^{(p)}\gamma_5) s 
+\overline s \epsilon_{\phi F}^{(p)} (A_{F21}^{(p)}+B_{F21}^{(p)}\gamma_5) d]
~~.
\label{equationc}
\end{equation}
\end{mathletters}
So for the amplitude $T$ for the $\Delta S=2$ process $s+s \to d +d$ we find, 
summing over the exchanges of Higgs eigenmodes $\epsilon_{\phi F}^{(p)}$ 
with squared masses $M_F^{2(p)}$, the formula 
(valid up to an overall phase)   
\begin{mathletters}
\label{allequations}
\begin{equation}
T=\sum_{p=\pm} \sum_{F=R,I} \overline d (A_{F12}^{(p)}+B_{F12}^{(p)}
\gamma_5) s 
{1 \over M_F^{2(p)}} \overline d (A_{F12}^{(p)}+B_{F12}^{(p)}\gamma_5) s ~~~,
\label{equationa}
\end{equation}
while from Sec. V and Table II we find for the squared masses, 
\begin{eqnarray}
M_R^{2(+)}=&&M_I^{2(-)}=(4 \lambda_{\phi}+2\mu_{1\phi}+2\mu_{2\phi}
+{7 \over 2} \alpha_{\phi}) \Omega_{\phi}^2~~~,\nonumber\\
M_R^{2(-)}=&&M_I^{2(+)}={9\over 2} \alpha_{\phi} \Omega_{\phi}^2,~~
\Omega_{\phi}^2={\lambda_{\phi} v_{\phi}^2 \over \lambda_{\phi}
-\mu_{1\phi}-\mu_{2\phi}- \alpha_{\phi}  }  ~~~.
\label{equationb}
\end{eqnarray}
\end{mathletters}
The Higgs exchange 
matrix element $\langle K|T| \overline K \rangle_{\rm Higgs}$ for 
the $ \overline K \to K$ transition, in the vacuum saturation approximation, 
then is given by 
\begin{mathletters}
\label{allequations}
\begin{equation}
|\langle K|T| \overline K \rangle_{\rm Higgs}| \simeq  N
|\langle K| \overline d \gamma_5 s|0 \rangle |^2 |D_{\rm Higgs}|~~~,
\label{equationa}
\end{equation}
with $D_{\rm Higgs}$ given by 
\begin{equation}
D_{\rm Higgs}=
\sum_{p=\pm}\sum_{F=R,I} {(B_{F12}^{(p)})^2 \over M_F^{2(s)} }~~~,
\label{equationb}
\end{equation}
\end{mathletters}
and with $N=8/3$ a Wick contraction and color factor.  

We wish now to compare the amplitude of Eq.~(47a) with the intermediate 
boson loop diagram contribution to the $K_L-K_S$ mass difference calculated 
by Gaillard and Lee [14], which in vacuum saturation approximation is in   
satisfactory agreement with experiment.  The Gaillard and Lee result
is 
\begin{mathletters}
\label{allequations}
\begin{equation}
|\langle K|T| \overline K \rangle_{\rm GL}| \simeq N 
|\langle  K|\overline d \gamma_{\mu} \gamma_5 s|0 \rangle |^2  
|D_{\rm GL}|~~~,
\label{equationa}
\end{equation}
with $|D_{\rm GL}|$ given by 
\begin{equation}
|D_{\rm GL}|={G_F^2 \over 4 \pi^2} M_c^2 s_{12}^2~~~,
\label{equationb}
\end{equation}
\end{mathletters}
in terms of the Fermi constant $G_F$, the charm quark mass $M_c$, and 
the sine of the Cabibbo angle $s_{12}=\sin \theta_C$.  To compare Eq.~(47a) 
to Eq.~(48a), we need the ratio of the pseudoscalar current to 
the axial vector 
current kaon to vacuum matrix elements, 
which can be estimated by standard current 
algebra methods (see, e.g., Shuryak [15]) to be 
\begin{equation}
{|\langle  K|\overline d \gamma_5 s|0 \rangle |^2   \over
|\langle  K|\overline d \gamma_{\mu} \gamma_5 s|0 \rangle |^2 } 
\simeq {\langle 0|\overline u u |0\rangle^2 \over M_K^2 f_K^4}
\simeq ({M_K \over M_s})^2 \simeq 11~~~,
\end{equation}
with $M_K$ and $f_K$ the kaon mass and decay constant and with 
$M_s$ the strange quark 
mass.  Combining everything, we find that the condition for the 
Higgs exchange contribution to the $K_L-K_S$ mass difference not to 
exceed the Gaillard and Lee estimate is 
\begin{equation}
|D_{\rm Higgs}| \leq {G_F^2 M_c^2 M_s^2 s_{12}^2 \over 
4 \pi^2 M_K^2}\simeq {2.6 \times 10^{-14} \over {\rm GeV}^2 }~~~,
\end{equation}
which will be used as the  strangeness changing neutral current constraint 
in the fits of the next section.  

\section*{ IX.~~~ Numerical Fits of the Three and Six Higgs Models
to the Experimental Data}

In order to fit the models to the experimental data, we follow the 
standard procedure of minimizing a ``cost function'' $C$, constructed as 
follows, 
\begin{mathletters}
\label{allequations}
\begin{equation}
C=C_{\rm mass}+C_{\rm CKM}+C_{\rm scnc}+C_{\rm parameter}~~~,
\label{equationa}
\end{equation}
with the pieces referring respectively to the constraints placed by 
fitting the masses, fitting the CKM mixing angles, obeying the strangeness 
changing neutral current bound of Eq.~(50), and keeping the asymmetry 
parameters as small as possible.  Before giving further details, we 
describe the general search method employed.  We perform all fits using 
the minimization routine ``powell'' of Press et. al. [12].  
As given in [12], this routine works well for the six Higgs model where the 
degeneracy between the first and second families is already broken, 
before inclusion of the asymmetry parameters, by 
the $\eta$ Higgs couplings.     
However, in the three Higgs model, it is not  a priori specified which states 
become the first and which become the second families, and so eigenvalue 
crossings can occur in the course of the iteration which result in 
discontinuous behavior of the cost function.  This causes a problem with 
the bracketing routine ``mnbrak'' of [12], which occasionally gets 
stuck in an indefinite loop.  The fix is simply putting an iteration 
counter into ``mnbrak'', to force an exit with a default bracketing 
(specifically, in terms of the quantities defined in ``mnbrak'', 
$c=a$, $fc=fa$) if convergence to a bracketing is not attained in 
$N_{\rm max}$ passes through the loop.  We found the same results in 
the three Higgs model with $N_{\rm max}=5$ as with $N_{\rm max}=30$, 
indicating that a bracketing is attained very rapidly, or not at all.  
As an additional check, we verified that the original and the modified 
versions of ``mnbrak'' give identical results for the six Higgs model, 
where level crossings and associated discontinuous behavior do not occur.

Let us now turn to the construction of the various cost function terms 
in Eq.~(51a), working throughout in units where 1 GeV$=$unity.  For the 
mass cost function, we use a standard chi squared function constructed 
from expected values of the masses and their estimated errors, including 
the electroweak mass parameter $v$ of Eq.~(33b).  To prevent the chi 
squared for certain very accurately known masses (such as the electron mass) 
from dominating the fits, 
we truncate these masses to a few significant figures and use enlarged  
error estimates.  In the six Higgs 
model we also add a term that 
favors fits with $\Omega_{\phi} \simeq \Omega_{\eta}$, since this degeneracy 
plays a role in the extension to neutrino mixings discussed in the next 
section; in practice, we find that this term has very little effect on   
the fits, since nearly equal values of $\Omega_{\phi}$ and $\Omega_{\eta}$   
are favored even in its absence.  (This term is omitted in the three Higgs 
model, where it is not relevant.)   Adding these contributions, we have 
for the mass cost function $C_{\rm mass}$, 
\begin{eqnarray}
C_{\rm mass}=&&\left({M_u-0.005 \over 0.003}\right)^2 
+\left({M_c-1.3 \over 0.18}\right)^2 
+\left({M_t-173. \over 6.}\right)^2 \nonumber\\
+&&\left({M_d -0.01 \over 0.005}\right)^2 + \left({M_s-0.2 \over 0.06}
\right)^2 
+\left({M_b-4.3 \over 0.2}\right)^2  \nonumber\\
+&&\left({M_e-0.00051 \over 0.0001}\right)^2 
+ \left({M_{\mu}-0.1057 \over 0.001}\right)^2
+\left({M_{\tau}-1.777 \over 0.001}\right)^2 \nonumber\\
+&&\left({ [6(\Omega_{\phi}^2 + \Omega_{\eta}^2)]^{1\over 2} -247. \over 3.}
\right)^2 \nonumber\\
+&&(\Omega_{\phi}-\Omega_{\eta})^2 ~~~.
\label{equationb}
\end{eqnarray}

For the strangeness changing neutral current cost function, we  use a 
chi squared function with expectation zero and standard deviation equal 
to the bound of Eq.~(50), 
\begin{equation}
C_{\rm scnc}=\left({ |D_{\rm Higgs}| \over 2.6 \times 10^{-14} }
\right)^2~~~.
\label{equationc}
\end{equation}
\end{mathletters}

To set up the CKM cost function, we make the standard rephasings to 
put the CKM matrix in the form 
\begin{mathletters}
\label{allequations}
\begin{equation}
U_{\rm CKM}=\pmatrix{1&s_{12}&s_{13}e^{-i\delta_{13}} \cr
                    -s_{12}&1&s_{23} \cr
                    -s_{13}e^{i\delta_{13}}& -s_{23}& 1 \cr}~~~,
\label{equationa}
\end{equation}
to first order accuracy in small quantities,  
and then construct a chi squared function from the expected 
values and estimated errors for $s_{12}$, $s_{13}$, and $s_{23}$.  
Although the CP violating angle $\delta_{13}$ has not been reliably 
determined experimentally, it appears likely that it is appreciable, so 
we also include a chi squared term requiring $|\sin \delta_{13}|$ to be 
equal to $0.6 \pm 0.3$, giving 
\begin{eqnarray}
C_{\rm CKM}=&&\left({s_{12}-0.221 \over 0.002}\right)^2 
+\left({s_{13}-0.0035 \over 0.0009}\right)^2 \nonumber\\
+&&\left({s_{23}-0.041\over 0.003}\right)^2
+\left({|\sin \delta_{13}|-0.6\over 0.3}\right)^2~~~.
\label{equationb}
\end{eqnarray}
\end{mathletters}
Altogether, then, there are 11 quantities to be fitted in $C_{\rm mass}$, 
1 to be fitted in $C_{\rm scnc}$, and 4 to be fitted in $C_{\rm CKM}$, for 
a total of 16.  

Let us now count the numbers of parameters in the two models, and establish   
the cost functions for the parameters.  
Despite its increased complexity in terms of particle content, the six 
Higgs model has the smaller number of parameters, since it violates CP 
only spontaneously and so all Yukawa couplings appearing in the Lagrangian 
are real. Altogether, there are 37 parameters that enter into the 
iterative fit  for the six Higgs model.  These are the $\phi$ and $\eta$ 
expectations $\Omega_{\phi}$ and $\Omega_{\eta}$, the real parts of the 
Yukawa couplings $g_{\phi,\eta}^f,~f=u,d,e$, the complex asymmetry 
parameters $\delta_{1,2}$ introduced in Eq.~(37), the angle $\theta$ of 
Eqs.~(21) and (39b), and the real asymmetry parameters $\beta_{\phi mn}^f,~
f=u,d,e,~m+n<6$ introduced in Eqs.~(6b, c). The parameters $\lambda_{\phi}$,  
$\mu_{1\phi}+\mu_{2\phi}$, and $\alpha_{\phi}$, which enter the calculation 
only through their appearance in the Higgs masses in the strangeness 
changing neutral current constraint [see Eqs.~(46-47)], were fixed at the 
respective values 1, 0.3, and 0.3, and were not iterated.  To 
construct the cost function for the iterated parameters, we note that no 
additional constraint is needed for the expectations $\Omega_{\phi,\eta}$ 
or the Yukawa couplings $g_{\phi,\eta}^f$ because these are already 
adequately controlled by $C_{\rm mass}$ of Eq.~(51b).  For the remaining 
parameters we use the cost function 
\begin{mathletters}
\label{allequations}
\begin{equation}
C_{\rm parameter}=\sum_{\matrix{n=1,2\cr}} \left|{\delta_n \over 
\sigma_{\rm parameter}}\right|^{\epsilon}
+\sum_{ \matrix{m,n~m+n<6\cr f=u,d,e\cr}} \left|{\beta_{\phi mn}^f 
\over \sigma_{\rm parameter}}\right|^{\epsilon} 
+ \left|{\theta \over 6.28} \right|^{\epsilon}~~~.
\label{equationa}
\end{equation}
In Eq.~(53a)  the exponent $\epsilon$ and the width $\sigma_{\rm parameter}$ 
are parameters of the fitting procedure, which effectively set up a model 
for how the small asymmetries are distributed.  We were able to get 
satisfactory fits for both $\epsilon=1$ and $\epsilon=2$, but convergence 
was much slower for the latter, suggesting that $\epsilon=1$ is a model in 
closer correspondence to the experimental data, and we shall only present 
the $\epsilon=1$ results in the discussion below.   
To initialize the six Higgs 
minimization search, we started from $\Omega_{\eta}=\Omega_{\phi}=70.7$,  
the values of Eq.~(34a) for the Yukawa couplings $g_{\phi,\eta}^f$, 
zero for the asymmetry  
parameters $\delta_n,\beta_{\phi mn}^f$, and zero for $\theta$.

Because the three Higgs model, to give a CP violating CKM matrix, must  
violate CP explicitly, its Yukawa couplings 
and Yukawa asymmetries can have imaginary parts, and so there are 
57 parameters that enter into the iterative fit.
These are the $\phi$ expectation $\Omega_{\phi}$, the real parts of the 
Yukawa couplings $g_{\phi}^f,~f=u,d,e$ and the imaginary part of 
$g_{\phi}^d$ (since $g_{\phi}^{u,e}$, which are not involved in the 
strangeness changing neutral current constraint, enter only through their 
absolute values, they can be rephased to be real), the complex asymmetry 
parameters $\delta_{1,2}$ introduced in Eq.~(36a),  
and the complex asymmetry parameters $\beta_{\phi mn}^f,~
f=u,d,e,~m+n<6$ introduced in Eqs.~(6b, c). Again, the parameters 
$\lambda_{\phi}$,  
$\mu_{1\phi}+\mu_{2\phi}$, and $\alpha_{\phi}$, which enter the calculation 
only through the strangeness 
changing neutral current constraint, were fixed at the 
respective values 1, 0.3, and 0.3.  To 
construct the cost function for the iterated parameters, we 
note that again no 
additional constraint is needed for the expectation $\Omega_{\phi}$ 
or the real parts of the Yukawa couplings $g_{\phi}^f$, because these are 
adequately controlled by $C_{\rm mass}$ of Eq.~(51b).  For the remaining 
parameters we use the cost function 
\begin{equation}
C_{\rm parameter}=\sum_{\matrix{n=1,2\cr F=R,I\cr}} 
\left|{\delta_{nF} \over \sigma_{\rm parameter}}\right|^{\epsilon}
+\sum_{\matrix{m,n~m+n<6\cr f=u,d,e\cr F=R,I\cr}} 
\left|{\beta_{\phi mn F}^f \over \sigma_{\rm 
parameter}}\right|^{\epsilon} 
+ \left|{g_{\phi I}^d \over 0.028}\right|^{\epsilon}~~~.
\label{equationb}
\end{equation}
\end{mathletters}
The width 0.028 governing $g_{\phi I}^d$ is chosen here as twice the 
natural magnitude of $g_{\phi}^d$ according to the estimate of Eq.~(34b), so 
as to bound $g_{\phi I}^d$ but not overly restrict it, much as the width for 
$\theta$ in the six Higgs model is chosen in Eq.~(53a) as twice the 
maximum magnitude 
$\pi$ of $|\theta|$.  Again, the exponent $\epsilon$ and the 
width $\sigma_{\rm parameter}$ 
are parameters that model how the small asymmetries are distributed.  For 
comparison with the six Higgs model fits, we shall again only present 
$\epsilon=1$ results in the discussion that follows.    
To initialize the three Higgs 
minimization search, we started from $\Omega_{\phi}=100$,  
the values of Eq.~(34b) for the real parts of the 
Yukawa couplings $g_{\phi}^f$,    
zero for the imaginary part of $g_{\phi}^d$,  and zero for the 
complex asymmetry parameters $\delta_n,\beta_{\phi mn}^f$.  

We begin by presenting results for the six Higgs model. In any fitting  
procedure involving more parameters than quantities to be fit, one has 
to worry about overfitting, and we deal with this in the following 
way.  As we shall see shortly, the most sensitive aspect of the fitting 
procedure for the six Higgs model is getting the CKM parameters correct, 
and so we take  
the cost function subcomponent $C_{\rm CKM}$ as a measure of overfitting.  
Making a series of fits using the cost function of Eq.~(53a) with 
$\epsilon=1$, as a function of the width $\sigma_{\rm parameter}$, we 
find that the value of $C_{\rm CKM}$ is a monotone decreasing function of 
the width.  For very small values of the width (i.e., asymmetries restricted 
to have very small values) we find a value of $C_{\rm CKM}$ much larger than 
4, the number of fitted CKM matrix degrees of freedom; for 
large values of the 
width we find values of $C_{\rm CKM}$ much less than 4, 
indicating overfitting.  
We take as ``good'' fits ones resulting from widths $\sigma_{\rm parameter}$ 
that yield a $C_{\rm CKM}$ of order 4; an example of such a fit, with 
$\sigma_{\rm parameter}=0.03$, is given in Table III.  This fit, which was  
attained after 229 iterations to achieve a one part in $10^6$ change in the 
cost function in an iteration (we will use this same convergence 
criterion throughout), had $C_{\rm mass}=0.13$, $C_{\rm CKM}=4.65$, 
$C_{\rm scnc}=3\times 10^{-4}$, and $C_{\rm parameter}=38.9$,  
giving a total cost function $C=43.7$. The 
values of the parameters giving this fit are as follows,
\begin{eqnarray}
&&\Omega_{\phi}=71.27,~ \Omega_{\eta}=71.27  ~~~,\nonumber\\
&&g_{\phi}^u=0.811,~g_{\phi}^d=0.0201,~g_{\phi}^e=0.00831~~~,\nonumber\\
&&g_{\eta}^u=0.00715,~g_{\eta}^d=0.00112,~g_{\eta}^e=0.000371~~~,\nonumber\\
&&\delta_{1R}=0.00269,~\delta_{2R}=0.0340,~\delta_{3R}=-0.0367~~~,\nonumber\\
&&\delta_{1I}=0.00074,~\delta_{2I}=0.0027,~\delta_{3I}=-0.0034~~~,\nonumber\\
&&\theta=150.8 \, {\rm degrees}~~~,\nonumber\\
&&\phantom{\vrule height 20pt}\nonumber\\
&&[\beta_{\phi}^u]=\pmatrix{0.1612    &0.0477      & 0.0268       \cr
                            0.0144    &0.00024     &-0.0133        \cr
                            -0.0442   &-0.0367     &-0.1562
\cr}~~~,\nonumber\\ 
&&\phantom{\vrule height 20pt}\nonumber\\
&&[\beta_{\phi}^d]=\pmatrix{0.1660    & 0.1589     &0.0189         \cr
                            0.        & 0.0180     &-0.0190        \cr
                            -0.1398   &-0.0189     &-0.1841
\cr}~~~,\nonumber\\ 
&&\phantom{\vrule height 20pt}\nonumber\\ 
&&[\beta_{\phi}^e]=\pmatrix{0.1038    & 0.         &-0.0517         \cr
                            0.        & -0.00081   &-0.0375         \cr
                            -0.0366   &-0.00011    &0.0230 \cr}~~~.
\end{eqnarray}

We see that the largest value of the $\beta$ asymmetry parameters is 
0.184 in magnitude, so the first question we must address is whether this 
large asymmetry is needed to reproduce the large mixing $s_{12}=0.221$ 
between the first and second families.  To show that this is not the case, 
we exhibit the result of rerunning the fit, this time omitting the $s_{13}$ 
and $s_{23}$ terms from the cost function.  The result, attained after 
137 iterations, has $C_{\rm mass}=0.04$ (that is, the fitted mass values 
are right on their targets) and $s_{12}=0.221$, so that the Cabibbo mixing 
is also right on target, but the largest of the $\beta$ 
asymmetry parameters has a magnitude of 0.01, a factor of 18 smaller than 
in the fit of Eq.~(54).  The values for the unconstrained third family  
mixings obtained this way are $s_{13}=0.00021,~s_{23}=0.00072$, much smaller 
than in the fit of Eq.~(54).  So we conclude that the large $\beta$ asymmetry  
values of Eq.~(54) are needed to get correct fits to the third family 
mixings; the correct value of $s_{12}$ by itself is obtained with $\beta$ 
values much smaller in magnitude than $s_{12}$, in agreement with our 
observation in Sec.~VII that $s_{12}$ is of zeroth order in the asymmetries.  

As a second experiment, which gives further insight into why the 
model requires large asymmetries to fit the third family mixings, 
we rerun the fit replacing the targets for both $s_{13}$ and $s_{23}$ by 
their geometric mean $\simeq 0.011$, with a standard deviation of 0.0015.  We 
find 
now convergence in 216 iterations, with $C_{\rm mass}=0.02$ 
(that is, again the 
fitted mass values are right on their targets), and values for the CKM 
mixings of $s_{12}=0.221$, $s_{13}=0.0117$, $s_{23}=0.0101$.  For the  other 
components of the cost function we find $C_{\rm CKM}=0.86$, $C_{\rm scnc} =
0.9\times 10^{-4}$, and $C_{\rm parameter}=3.5$, for a total of $C=4.4$. 
As suggested by the small value of $C_{\rm parameter}$, the largest  
of the $\beta$ asymmetry parameters now has a magnitude of 0.028, a factor of 
6.6 smaller than in the fit of Eq.~(54).  We conclude from this fit that 
what requires the large asymmetries in Eq.~(54) is splitting $s_{23}$ 
and $s_{13}$ from a common mean value.  

This conclusion can be understood from a simple analytic model, in   
which the corrections of Appendix C to $U_L^f$ and $U_R^{f\dagger}$ 
are neglected.  
Referring to Eq.~(42a), let us write $V_{\rm CKM}$ to first order
accuracy as  
\begin{mathletters}
\label{allequations}
\begin{equation}
V_{\rm CKM}=\pmatrix{ 1 &v_{12} \cr -v_{12}^* & 1 \cr }  ~~~,
\label{equationa}
\end{equation}
so that the fitted $s_{12}$ is given by $s_{12}=|v_{12}|$.  Then from the 
approximation of Eq.~(42b) for $U_{\rm CKM}$, together with the 
CP invariance condition [see Eq.~(38b) and the discussion
following Eq.~(39c)] $\sigma_{13}^f=\sigma_{23}^{f*}$, we find that  
\begin{eqnarray}
s_{13}=&&|s_3-d_3|/3,~ s_{23}=|s_3+d_3|/3~~~,\nonumber\\
s_3\equiv &&\sigma_{13}^u-\sigma_{13}^d,~  
d_3\equiv v_{12} \sigma_{23}^d~~~.
\label{equationb}
\end{eqnarray}
\end{mathletters}
Thus, the spread of $s_{13}$ and $s_{23}$ from their geometric mean 
is governed by $d_3$, in which the quantity $\sigma_{23}^d$, which is 
a linear combination of the $\beta$ asymmetries, is suppressed in magnitude 
by a factor of $|v_{12}|=s_{12}=0.221$.  This is why large $\beta$ 
asymmetries are needed to fit the experimental data, whereas much smaller 
asymmetries suffice when the observed $s_{13}$ and $s_{23}$ are replaced in 
the fitting program by their geometric mean.  For example, 
in the fit of Eq.~(54), the  magnitude of $d_3$ is 0.0445, which 
corresponds to a value $\sigma_{23}^d=0.0445/0.221=0.20$, 
similar in size to the  maximum $\beta$ asymmetries found in the fits.  
Thus, the six Higgs model interprets the large difference in magnitude 
between the observed $s_{13}$ and $s_{23}$ as indicating asymmetries 
in the Yukawa couplings  
substantially larger than one might naively infer from the magnitude 
of $s_{23}$.  The possible relevance of this observation to the extension  
of our model to neutrino mixing will be discussed in Sec.~X.  
 
We next address issues of fine tuning and naturalness in the six Higgs 
model.  In the fit of Eq.~(54), the absolute values of the matrix 
elements of the matrices $U_L^{u,d}$ and $U_R^{d\dagger}$ take the values
\begin{mathletters}
\label{allequations}
\begin{eqnarray}
[|U_L^u|]=\pmatrix{0.974 & 0.224 & 0.055\cr
                 0.224 & 0.974 & 0.034\cr
                 0.046 & 0.045 & 1.000 \cr }~~~,
\label{equationa}
\end{eqnarray}
			
\begin{equation}
[|U_L^d|]=\pmatrix{1.000 & 0.00010 & 0.066 \cr  
                 0.00010 & 1.000 & 0.067 \cr
                 0.066 & 0.067 & 1.000\cr }~~~,
\label{equationb}
\end{equation}
\begin{equation}
[|U_R^{d\dagger}|]=\pmatrix{1.000 & 0.00010 & 0.033 \cr  
                 0.00010 & 1.000 & 0.036 \cr
                 0.033 & 0.036 & 1.000\cr }~~~.
\label{equationc}
\end{equation}
\end{mathletters}
We see that the mixing $s_{12}$ of the first two families  
arises nearly entirely from $U_L^u$, while the 
$2 \times 2$ submatrices of $U_L^d$ and $U_R^{d\dagger}$,
that mix the first two families (and that are equal to good accuracy) are 
nearly the unit matrix, which is what allows the strangeness changing 
neutral current constraint to be satisfied.  To estimate the amount of 
fine tuning involved in this, we note that $|D_{\rm Higgs}|$ of Eq.~(47b) 
is quadratic in the matrix 
element $|U_{L12}^d| \simeq |U_{R12}^{d\dagger}|$.   
Hence if the entry for $|D_{\rm Higgs}|/(2.6 \times 10^{-14})$ in Table III 
were scaled up from 0.016 to unity, corresponding to the strangeness 
changing neutral current constraint being just 
barely satisfied, the off diagonal matrix elements 
$|U_{L12}^d| \simeq |U_{R12}^{d\dagger}|$ in Eqs.~(56b, c) 
would be scaled up 
from 0.00010 to $0.00010/0.016^{1\over 2} \simeq 0.00079$.   
Taking as a ``generic'' off 
diagonal matrix element the average value $\simeq 0.05$ of the 13 and 23 
matrix elements of Eqs.~(56b, c), we estimate that fine tuning in the mixing 
matrices, of order a factor of $0.05/0.00079 \simeq 63$, 
is involved in satisfying the 
strangeness changing neutral current constraint, 
for an assumed Higgs mass in the  fit [see the second line in Eq.~(46b)] 
of $M_R^{(-)} = (4.5 \times 0.3)^{1\over 2} \Omega_{\phi} \simeq 83$ GeV. For 
a Higgs mass of 330 GeV  the fine tuning would be correspondingly 
reduced to a factor of roughly 16, and for a Higgs mass of 800 GeV the   
fine tuning factor would be roughly 6.  

Given that there is some fine tuning involved in obeying the strangeness 
changing neutral current constraint, one can ask whether it is natural or 
unnatural to the experimental data.  If the fine tuning is not natural 
to the data being fit, one would expect the fits to the masses and CKM 
parameters to improve, or the convergence to a fit to become faster, when 
the cost function term $C_{\rm scnc}$ is omitted from the total cost 
function.  Performing this experiment, we find that without $C_{\rm scnc}$, 
a comparably good fit is obtained ($C_{\rm mass}=0.64$, $C_{\rm CKM}=4.1$) 
as with the cost function term $C_{\rm scnc}$ included, but 
600 iterations, as opposed to 229, are required for comparable convergence.  
In other words, the strangeness changing neutral current constraint appears 
to guide the search to a region of parameter space that gives a good fit; 
we interpret this as an indication that the fine tuning involved 
in satisfying 
this constraint is in fact natural to the data.  
 
One other place where there is fine tuning in the fits is in the first 
family masses, since these are naturally zero only in the absence of 
Yukawa coupling asymmetries.  In principle, if the first family  
cost function terms are omitted from $C_{\rm mass}$, one might expect 
first family masses as large as 0.2  (the value of the maximum asymmetry 
parameters) times the corresponding third family masses,  which would 
give $M_u\sim 35,~ M_d\sim 0.9 ,~  M_e \sim 0.4$.  However, performing the 
experiment of omitting first family mass constraints from the fit, 
we find first family masses $M_u \simeq 0.9,~ M_d \simeq 0.23~, 
M_e \simeq 0.07$, that is, the first family masses are still smaller (or 
in the case of $M_d$, equal to) the second family masses.  We interpret 
this as an indication that small first family masses are in fact natural 
to the remaining experimental data when first family masses are excluded, 
in the framework of the six Higgs model.  

We conclude this section by giving some comparative fits in the three Higgs 
model.  Using the same cost function parameters and convergence criterion 
as in the six Higgs case, we get the three Higgs model fit shown in  
Table IV, which required 864 iterations.  
The mass fit is generally good, except for the low value 
$M_s=0.037$ (corresponding to $C_{\rm mass}=7.3$), while the CKM 
parameters are close to their targets ($C_{\rm CKM}=0.5$).  
When the strangeness 
changing neutral current constraint is omitted in this case,  we find 
faster convergence (398 iterations) and a better fit, with $M_s=0.151$ 
(corresponding to $C_{\rm mass}=0.7$) and with the CKM parameters 
right on target ($C_{\rm CKM}=0.1$).  This behavior contrasts sharply with 
what we saw in the six Higgs model fits, and we interpret 
it as indicating that 
the strangeness changing neutral current constraint is not natural to the 
data as interpreted in the three Higgs model. 

\section*{ X.~~~Experimental Issues, Neutrino Mixing, Coupling Constant
Unification, and Directions for Future Work}

Of the two models that we have developed in the previous sections, we 
find the six Higgs model the more interesting as a candidate for an  
extension of the standard model into the energy region that will become   
accessible in the next decade.  As compared with the three Higgs model, 
the six Higgs model has fewer parameters, gives better overall fits to the 
data, and gives some indication that the strangeness changing neutral  
current constraint is natural to the data.  It also violates CP spontaneously 
in an interesting way that is correlated with the generation of second 
family masses for the $u,d,e$ families.  

The prime experimental signature of the six Higgs model is the spectrum 
of Higgs states tabulated in Table II.  If the potential $V_2$ that 
couples the $\phi$ to $\eta$ Higgs overall phases is in fact small, 
then the lightest Higgs states should be the pseudo-Goldstone bosons.  
However, because of the ${1\over 2}$ power scaling law of Eq.~(25c), they 
need not be so light as to conflict with current Higgs mass limits.  For 
example, if the Higgs masses $M_{\rm Higgs}$ that enter into the 
strangeness changing neutral current constraint are of order 330 GeV, and 
$V_2/V_{\phi,\eta} \sim 0.1$, which is in the weak coupling regime, 
then the pseudo-Goldstone boson masses 
are expected to be of order $(0.1) ^{1 \over 2} 330$ GeV $\simeq 104$ GeV, 
above current Higgs mass limits.  

Although we have included the possibility of a right handed neutrino, and of
Dirac neutrino masses and mixing analogous to CKM mixing, in our Lagrangian, 
we have not attempted a detailed study of the neutrino sector because the 
experimental picture there is still incomplete.  However, let us briefly  
address the recent report by the Super-Kamiokande Collaboration [16] 
of evidence for atmospheric neutrino oscillations, 
suggesting large mixing  
(of order unity) of second and third family neutrinos.  This is   
clearly a different 
pattern than is seen for the charged fermion mixings, where, for    
example, in the fit of Eq.~(54) the $\mu - \tau$ mixing matrix elements of 
$U_L^e$ are smaller than 0.01 in magnitude.  Large $\nu_{\mu} - \nu_{\tau}$ 
mixing can be accommodated                                      
in our model, nonetheless, by assuming that the Yukawa coupling ratio  
$g_{\eta}^f/g_{\phi}^f$, which we have taken to be small for $f=u,d,e$, is 
close to unity for $f=\nu$.  Together with $\Omega_{\eta}/\Omega_{\phi}  
\simeq 1$, this implies that $R^{\nu}$ of Eq.~(39b) is close to unity in 
magnitude (although it can have a nonzero phase).  
Referring to Eqs.~(38a, b),  we see that this implies that the neutrino 
mass matrix is now nearly degenerate in the two dimensional subspace spanning 
the 
second and third families, and so small asymmetries, or asymmetries nearly 
equal in magnitude, then imply nearly 
maximal mixing.  To show this explicitly, let us apply the analysis of  
Appendix B to the mass matrix 
\begin{mathletters}
\label{allequations}
\begin{equation}
m=\pmatrix{R^{\nu} & {1 \over 3} \sigma_{23} \cr
         {1 \over 3} \sigma_{32} & 1 \cr } ~~~. 
\label{equationa}
\end{equation}
Then for $M_L=mm^{\dagger}$, we have from Eq.~(B2b), 
\begin{equation}
M_L=\pmatrix{ |R^{\nu}|^2+{1\over 9}|\sigma_{23}|^2 &   
                 {1\over 3}(R^{\nu}\sigma_{32}^*+\sigma_{23})\cr
                  {1\over 3}(\sigma_{32} R^{\nu\,*}+\sigma_{23}^*) & 
                  1+{1\over 9}|\sigma_{32}|^2 \cr}~~~,
\label{equationb}
\end{equation}
and so Eq.~(B4b) gives for the mixing angle 
\begin{equation}
\Theta= {1\over 2}\tan^{-1} \left({ -{2 \over 3} 
|R^{\nu}\sigma_{32}^*+\sigma_{23}|   
\over 
|R^{\nu}|^2-1+{1\over 9}(|\sigma_{23}|^2 -|\sigma_{32}|^2)}  
\right) ~~~.
\label{equationc}
\end{equation}
Thus there is maximal mixing whenever 
\begin{equation}
{2\over 3} |R^{\nu}\sigma_{32}^*+\sigma_{23}| >>  
|R^{\nu}|^2-1+{1\over 9}(|\sigma_{23}|^2 -|\sigma_{32}|^2)~~~.
\label{equationd}
\end{equation}
\end{mathletters}
If $|R^{\nu}|$ is close to unity, this inequality can be satisfied either 
(i) if $\sigma_{23}$ and $\sigma_{32}$ are both small, or (ii) if the 
magnitudes of $\sigma_{23}$ and $\sigma_{32}$ are not small, but are 
approximately equal.
In Sec. IX we saw that to reproduce the observed 
CKM parameters $s_{23}$ and $s_{13}$, we needed sizable asymmetries (of order 
0.2), which if also present in the neutrino sector $\beta$'s would allow 
near maximal mixing of the second and third family neutrinos by case (ii) 
even when the ratio $R_{\nu}$ is only approximately unity in magnitude.  
Thus large $\nu_{\mu}-\nu_{\tau}$ mixing is easy to achieve in the six 
Higgs model.  Less natural is near degeneracy of the masses of $\nu_e$ 
and $\nu_{\mu}$, as appears to be needed  for both the 
MSW and the vacuum oscillation interpretations of the solar 
neutrino data, since the first family masses are zero in our model in the   
absence of Yukawa asymmetries.  
Such a degeneracy would have to be the result of sizable 
asymmetries  together with  
substantial fine tuning in the neutrino mass matrix, either to raise the  
$\nu_e$ mass to close to the $\nu_{\mu}$ mass in case (i), or to lower 
the $\nu_{\mu}$ mass to close to zero in case (ii) (as, for example,  
is done in the  model of Barger, Pakvasa, Weiler, and Whisnant [17];
see also Baltz, Goldhaber, and Goldhaber [17]).  
In either case, there will almost certainly be large mixing of $\nu_e$  
with $\nu_{\mu}$; so on 
this (very preliminary) interpretation, our model would favor the 
large angle as opposed to the small angle MSW solution.

Let us next address the issue of coupling constant unification in the six 
Higgs model.  Because we do not alter the fermion representation content of 
the standard model, the usual running coupling analysis applies.    
As noted by Langacker [18], the standard model with $\simeq$ 7 (by current   
data [19], 7.66)  Higgs boson 
doublets gives one loop coupling constant unification with a 
unification energy of order $5 \times 10^{13}$ GeV.  
Even with only 6 Higgs doublets, 
the magnitude of two loop 
radiative corrections [20] is sufficient to make coupling constant 
unification a possibility.  Of course, because the unification energy  
is lower than in the customary scenario, a mechanism is needed to 
suppress proton decay, such as is present in the  
$SU(15)$ family [21] of grand unification models.  Clearly, definitive  
statements here will depend on the nature of the high energy theory for 
which the six Higgs model is a low energy effective theory; the point 
we wish to stress, though, is that the six Higgs model may be a candidate 
for coupling constant unification without the assumption of low 
energy supersymmetry.  Whether such a candidate is needed, 
of course, will depend  
on the outcome of supersymmetry searches over the next decade.  

There are a number of obvious directions for further work on the models 
we have developed in this paper.  Entirely within the low energy effective 
action framework, one can address the issue of one loop radiative corrections 
to the mass and mixing matrix analysis given here.  This will involve the 
parameters determining the Higgs masses in an integral way, and  
if the six Higgs model is to be viable, the one loop corrections 
should improve, rather than make worse,  
the comparisons with experiment and the consistency tests discussed in 
Sec. IX. Another issue that can be addressed within the low energy 
framework is the magnitude of 
electroweak baryogenesis in the six Higgs model, and 
cosmological implications of this model more generally.  
At a deeper level, there is the issue of finding 
a grand unified model,  
composite model, or hybrid model comprising elements of both, which is 
a natural high energy physics source for the low energy effective 
action physics 
described by the six Higgs model. Such a high energy model must, through its 
representation content and 
instanton physics, justify the discrete chiral transformation rules assumed 
in Eqs.~(1a-c), and it is also the place where one must seek explanations 
for the ``vertical''  hierarchy of Yukawa coupling strengths, and the 
pattern of Yukawa coupling asymmetries, that is needed for our fits.    

\acknowledgments
This work was supported in part by the Department of Energy under
Grant \#DE--FG02--90ER40542.  I wish to thank Henry Frisch, Harald Fritzsch, 
Chris Kolda, Burt Ovrut, 
Jon Rosner, Bill Scott, and Sam Treiman for stimulating conversations or   
correspondence.   I also wish to acknowledge the hospitality of the 
Aspen Center for Physics, where the manuscript was completed.  

{\it Added note.}  After this paper was posted to the Los Alamos e-print 
archive, two earlier papers that use families of Higgs scalars (although 
without the ingredient of $Z_6$ discrete chiral symmetry analyzed here) were 
brought to my attention.  The paper of Derman and Jones [22] studies  
a two family, two Higgs doublet model with an $S_2$ permutation symmetry, and 
is probably the earliest paper to extend the idea of family symmetries 
to the Higgs sector; the paper of Derman [23] extends this to three 
families of fermions and Higgs doublets with an $S_3$ permutation symmetry.  

\appendix
\section{Numerical Minimization of the Higgs Potential}

Because the Higgs potential of Eqs.~(7a-c) is complicated, even with the 
simplifying assumptions of CP invariance and cyclic permutation symmetry, 
we have 
supplemented our analytic studies of the Higgs extrema with numerical 
studies, performed by using the conjugate gradient method to minimize 
the Higgs potential.  Since it is easy to analytically compute the 
first derivatives (the gradients) of the Higgs potential, 
it is advantageous to use 
the conjugate gradient method in a form where both the function to  
be minimized and its derivatives are externally supplied; this gives 
a faster routine and there is some built in redundancy that serves as a 
check, since the same information is in effect furnished twice, once through 
the computation of the function and a second time through the 
independent computation 
of its derivatives.  We have used the minimization program ``frprmn'' of 
Press et al. [12], with the following  modification.  
Press et al.   
base the convergence criterion in their program on computing the change 
in the {\it function value} over one iteration, but this results in 
significant truncation error inaccuracies for the minimizing values of 
the {\it arguments} (the Higgs fields) when the function is large 
in magnitude but very flat at its minimum.  Since the gradients are 
explicitly known, and since at the minimum the gradients must all 
vanish, much better accuracy for the minimizing Higgs fields 
is obtained by making the convergence criterion depend on the maximum 
gradient. With this modification to ``frprmn'', one can verify vanishing  
of the gradients to double precision accuracy at the Higgs potential 
minimum.  

To obtain the formulas for the gradients, as a function of general 
$\phi,\eta$, we substitute $\phi \to \phi + \delta \phi~, 
\eta \to \eta + \delta \eta$ into Eqs.~(7a-c), and retain the first 
order variations, which can be brought to the convenient form 
\begin{mathletters}
\label{allequations}
\begin{equation}
\delta {\cal L}_{\rm Higgs~potential}
={\rm Re}\sum_{n=1}^3 (C_n^{\phi}\delta \phi_n^* + C_n^{\eta}\delta \eta_n^*)
~~~.
\label{equationa}
\end{equation}
We assume both CP invariance and cyclic permutation symmetry; using the 
latter we get formulas for $C_{2,3}^{\phi,\eta}$ by cyclic permutation of 
the arguments of $C_1^{\phi,\eta}$.  Changing notation for the 
coefficients from  
$C_{\ell mn}$ to $C_{\ell;mn}$, to avoid notational ambiguities 
when explicit numerical values are assigned for $m$,  
we obtain the following explicit expressions for $C_1^{\phi,\eta}$. 
\begin{eqnarray}
C_1^{\phi}=&&4 \lambda_{\phi}(\phi_1^*\phi_1-v_{\phi}^2) \phi_1
-2(\mu_{1\phi}+\mu_{2\phi})(\phi_2^*\phi_2+\phi_3^*\phi_3)\phi_1 \nonumber\\
-&&\alpha_{\phi}(2\phi_2\phi_1^*\phi_3+\phi_3^2\phi_2^*+\phi_2^2\phi_3^*) 
+ \gamma \eta_1 \nonumber\\
+&&\sum_{m=1}^3 [2C_{1;1m} \eta_m^*\eta_m\phi_1 \nonumber\\
+&&C_{2;1m} \eta_1\eta_m^*\phi_m + C_{2;m1}\eta_1\eta_m^*\phi_m\nonumber\\
+&&C_{3;1m}\phi_2\eta_m^*\eta_{m-1}+C_{3;3m}\phi_3\eta_{m-1}^*\eta_m\nonumber\\
+&&C_{4;m1}\phi_3\eta_m^*\eta_{m+1}+C_{4;m2}\phi_2\eta_m\eta_{m+1}^*\nonumber\\
+&&C_{5;1m}\eta_2\eta_m^*\phi_{m-1}+C_{5;m2}\eta_2\eta_{m+1}^*\phi_m\nonumber\\
+&&C_{6;m1}\eta_3\eta_m^*\phi_{m+1}+C_{6;3m}\eta_3\eta_{m-1}^*\phi_m\nonumber\\
+&&C_{7;1m}\phi_2\phi_m^*\eta_{m-1}+C_{7;3m}\phi_3\phi_m\eta_{m-1}^*
+C_{7;m1}\phi_m^*\phi_{m+1}\eta_3\nonumber\\
+&&C_{8;1m}\phi_2\eta_m^*\phi_{m-1}+C_{8;3m}\phi_3\eta_m\phi_{m-1}^*
+C_{8;m2}\phi_m\phi_{m+1}^*\eta_2\nonumber\\
+&&C_{9;m1}\eta_m^*\eta_{m+1}\eta_3\nonumber\\
+&&C_{10;m2}\eta_m\eta_{m+1}^*\eta_2\nonumber\\
+&&C_{11;1m}\eta_2\phi_m^*\eta_{m-1}+C_{11;m1}\phi_m^*\eta_{m+1}
\eta_3\nonumber\\
+&&C_{12;3m}\eta_3\eta_m\phi_{m-1}^*+C_{12;m2}\eta_m\phi_{m+1}^*\eta_2]~~~,
\label{equationb}
\end{eqnarray}
\vspace{20pt}
\begin{eqnarray}
C_1^{\eta}=&&4 \lambda_{\eta}(\eta_1^*\eta_1-v_{\eta}^2) \eta_1
-2(\mu_{1\eta}+\mu_{2\eta})(\eta_2^*\eta_2+\eta_3^*\eta_3)\eta_1 \nonumber\\
-&&\alpha_{\eta}(2\eta_2\eta_1^*\eta_3+\eta_3^2\eta_2^*+\eta_2^2\eta_3^*)
+\gamma \phi_1 \nonumber\\
+&&\sum_{m=1}^3 [2C_{1;m1} \phi_m^*\phi_m\eta_1 \nonumber\\
+&&C_{2;m1}\phi_1\phi_m^*\eta_m+C_{2;1m}\phi_1\phi_m^*\eta_m\nonumber\\
+&&C_{3;m1}\eta_3\phi_m^*\phi_{m+1}+C_{3;m2}\eta_2\phi_m\phi_{m+1}^*\nonumber\\
+&&C_{4;1m}\eta_2\phi_m^*\phi_{m-1}+C_{4;3m}\eta_3\phi_m\phi_{m-1}^*\nonumber\\
+&&C_{5;m1}\phi_3\phi_m^*\eta_{m+1}+C_{5;3m}\phi_3\phi_{m-1}^*\eta_m\nonumber\\
+&&C_{6;1m}\phi_2\phi_m^*\eta_{m-1}+C_{6;m2}\phi_2\phi_{m+1}^*\eta_m\nonumber\\
+&&C_{7;m2}\phi_m\phi_{m+1}^*\phi_2\nonumber\\
+&&C_{8;m1}\phi_m^*\phi_{m+1}\phi_3\nonumber\\
+&&C_{9;1m}\eta_2\phi_m^*\eta_{m-1}+C_{9;3m}\eta_3\phi_m\eta_{m-1}^*
+C_{9;m2}\eta_m\eta_{m+1}^*\phi_2\nonumber\\
+&&C_{10;1m}\eta_2\eta_m^*\phi_{m-1}+C_{10;3m}\eta_3\eta_m\phi_{m-1}^*
+C_{10;m1}\eta_m^*\eta_{m+1}\phi_3\nonumber\\
+&&C_{11;3m}\phi_3\phi_m\eta_{m-1}^*+C_{11;m2}\phi_m\eta_{m+1}^*
\phi_2\nonumber\\
+&&C_{12;1m}\phi_2\eta_m^*\phi_{m-1}+C_{12;m1}\eta_m^*\phi_{m+1}
\phi_3]~~~.\nonumber\\
\label{equationc}
\end{eqnarray}
\end{mathletters}

\section{Bi-Unitary Diagonalization of a $2\times 2$ Matrix}

We give here the method for constructing the matrices $V_L$ and 
$V_R^{\dagger}$ that obey Eq.~(41a) of the text, suppressing the flavor 
index $f$ throughout.  Let $m$ by the $2\times 2$ complex matrix defined by 
\begin{equation}
$$m=\pmatrix{\sigma_{11} & \sigma_{12} \cr \sigma_{21} & \sigma_{22} \cr 
}~~~. 
\end{equation}
We begin by forming the self-adjoint matrices $M_L\equiv mm^{\dagger}$ and 
$M_R \equiv m^{\dagger}m$, which we write in the form 
\begin{mathletters}
\label{allequations}
\begin{equation}
M_L=\pmatrix{A_L & z_L^* \cr z_L & B_L \cr },~~~ 
M_R= \pmatrix{A_R & z_R^* \cr z_R & B_R \cr }~~~,
\label{equationa}
\end{equation}
with 
\begin{eqnarray}
A_L=&&|\sigma_{11}|^2+|\sigma_{12}|^2~~~,\nonumber\\
B_L=&&|\sigma_{21}|^2+|\sigma_{22}|^2~~~,\nonumber\\
z_L=&&\sigma_{11}^*\sigma_{21}+\sigma_{12}^*\sigma_{22}~~~,
\label{equationb}
\end{eqnarray}
and 
\begin{eqnarray}
A_R=&&|\sigma_{11}|^2+|\sigma_{21}|^2~~~,\nonumber\\
B_R=&&|\sigma_{12}|^2+|\sigma_{22}|^2~~~,\nonumber\\
z_R=&&\sigma_{12}^*\sigma_{11}+\sigma_{22}^*\sigma_{21}~~~.
\label{equationc}
\end{eqnarray}
The quantities just defined are not independent, since it is easy to verify 
that 
\begin{eqnarray}
A_L+B_L=&&A_R+B_R ~~~,\nonumber\\
{1\over 4} (A_L-B_L)^2+|z_L|^2=&&{1\over 4} (A_R-B_R)^2+|z_R|^2~~~,\nonumber\\
|z_L|^2 \leq A_L B_L,&&~~~|z_R|^2 \leq A_R B_R~~~.
\label{equationd}
\end{eqnarray}
\end{mathletters}
The desired bi-unitary matrices will be the $V_L$ for which 
$V_LM_LV_L^{\dagger}$ is diagonal, and the $V_R$ for which 
$V_RM_RV_R^{\dagger}$ is diagonal, with eigenvalues ordered in magnitude.   

Thus, defining the self-adjoint matrix $M$ by 
\begin{mathletters}
\label{allequations}
\begin{equation}
M=\pmatrix{A& z^* \cr z & B \cr },~|z|^2\leq AB~~~,
\label{equationa}
\end{equation}
it suffices to find the diagonalizing unitary transformation $V$ 
that yields 
\begin{equation}
V M V^{\dagger}= \pmatrix{ |\kappa_1|^2 & 0\cr 0 & |\kappa_2|^2 \cr }
,~~|\kappa_1| \leq |\kappa_2|~~~; 
\label{equationb}
\end{equation}
then all that we have to do is to apply this construction twice, first to 
$M_L$ and then to $M_R$.  Let us write  $M$ in Pauli matrix form as 
\begin{equation}
M={1\over 2}(A+B) + \vec v \cdot \vec \tau,~
\vec v=\left(z_R,z_I,{1\over 2}(A-B)\right)~~~,
\label{equationc}
\end{equation}
\end{mathletters}
with $z_{R,I}$ the real and imaginary parts of $z$.  Representing 
the diagonalizing $V$ in Pauli matrix form as 
\begin{mathletters}
\label{allequations}
\begin{equation}
V=\exp(i\Theta \vec n \cdot \vec \tau)=\cos \Theta
+ i \vec n \cdot \vec \tau  \sin \Theta~~~,
\label{equationa}
\end{equation}
and letting $\hat z =(0,0,1)$  be the unit vector in the third axis 
direction, 
a simple calculation shows that we satisfy Eq.~(B3b) by taking  
\begin{eqnarray}
\sin 2\Theta =&&{|\hat z \times \vec v|\over|\vec v|}
={|z|\over [{1\over 4} (A-B)^2 +|z|^2]^{1 \over 2} }~~~,\nonumber\\
\cos 2\Theta =&&{-\hat z \cdot \vec v\over|\vec v|}
={-{1 \over 2}(A-B) \over[{1\over 4} (A-B)^2 +|z|^2]^{1 \over 2}
}~~~,\nonumber\\ 
\Theta=&&{1\over 2} \tan^{-1}\left(-2|z| \over A-B \right)~~~,\nonumber\\
\hat n=&&-{\hat z \times \vec v \over |\hat z \times \vec v| }
={(z_I,-z_R,0)\over|z|}~~~,
\label{equationb}
\end{eqnarray}
and that this $V$ gives  
\begin{equation}
V M V^{\dagger} ={1\over 2} (A+B) -|\vec v| \tau_3~~~.
\label{equationc}
\end{equation}
\end{mathletters}
Thus we see that the squared eigenvalues are 
\begin{mathletters}
\label{allequations}
\begin{equation}
|\kappa_1|^2={1 \over 2} (A+B)-|\vec v|,~~ 
  |\kappa_2|^2={1 \over 2} (A+B)+|\vec v|~~~,
\label{equationa}
\end{equation}
which are correctly ordered; the smaller squared eigenvalue 
is guaranteed to be nonnegative by virtue of the fact that the product of  
the squared eigenvalues is 
\begin{equation}
{1\over 4} (A+B)^2 -|\vec v|^2 =AB-|z|^2 \geq 0~~~.
\label{equationb}
\end{equation}
\end{mathletters}
When $|z|=0$, the above formulas are indeterminate; we then get the correct 
eigenvalue ordering by taking $\sin 2\Theta=0$ 
and $\cos 2\Theta=\pm1$, with the $+$ sign holding for $A\leq B$ and the 
$-$ sign holding for $A > B$.  Referring back to the identities of 
Eq.~(B2d), we see that they imply that $|\vec v_L|=|\vec v_R|$, and thus 
the eigenvalues are the same 
for $M_L$ and $M_R$, as expected.   Substituting the expression for 
$\hat n$ in Eq.~(B4b) back into Eq.~(B4a), we get the further useful 
expression 
\begin{equation}
V=\pmatrix{ \cos \Theta & - {z^* \over |z|} \sin \Theta \cr
             {z \over |z|} \sin \Theta & \cos \Theta  \cr }
             ~~~.
\end{equation}

\section{Improved Formulas for $U_L^f$ and $U_R^{f\dagger}$}

In our numerical work, we used an improved approximation to $U_L^f$ 
and $U_R^{f\dagger}$ obtained by adding to Eqs.~(41b, c) the respective 
corrections $\Delta U_L^f$ and $\Delta U_R^{f\dagger}$, given by
\begin{mathletters}
\label{allequations}
\begin{equation}
\Delta U_L^f=\pmatrix{0&0&0&\cr 0&0&-{1\over 9} \kappa_2^f 
\eta_{32}^{f*}\cr                
{1\over 9} \kappa_2^{f*}\eta_{32}^f V_{L21}^f & 
{1\over 9} \kappa_2^{f*}\eta_{32}^f V_{L22}^f&0\cr }~~~,
\label{equationa}
\end{equation}
and 
\begin{equation}
\Delta U_R^{f\dagger}=\pmatrix{0&0& 
{1 \over 9} V_{R12}^{f\dagger} \kappa_2^{f*}\eta_{23}^f \cr 
0&0&{1 \over 9} V_{R22}^{f\dagger} \kappa_2^{f*}\eta_{23}^f \cr  
0&-{1\over9} \kappa_2^f \eta_{23}^{f*} & 0\cr}~~~.
\label{equationb}
\end{equation}
\end{mathletters}
Here $V_{L,R}^f$ are the matrices defined in Eq.~(41a) and computed in 
Appendix B, $\kappa_2^f$ is the eigenvalue defined in Eq.~(41a), given 
explicitly by 
\begin{mathletters}
\label{allequations}
\begin{equation}
\kappa_2^f=
V_{L21}^f(\sigma_{11}^f V_{R12}^{f\dagger}+\sigma_{12}^fV_{R22}^{f\dagger})+  
V_{L22}^f(\sigma_{21}^f V_{R12}^{f\dagger}+\sigma_{22}^fV_{R22}^{f\dagger})  
~~~,
\label{equationa}
\end{equation}
and the quantities $\eta^f_{23}$, $\eta^f_{32}$ are defined by 
\begin{eqnarray}
\eta_{23}^f=&&V_{L21}^f \sigma_{13}^f+V_{L22}^f \sigma_{23}^f~~~,\nonumber\\
\eta_{32}^f=&&\sigma_{31}^fV_{R12}^{f\dagger} + \sigma_{32}^f 
V_{R22}^{f\dagger}~~~.
\label{equationb}
\end{eqnarray}
\end{mathletters}
These corrections make the formulas for $U_L^f$ and $U_R^{f \dagger}$ 
accurate to first order when $(|\kappa_2^f|/3)^2$, rather 
than $|\kappa_2^f|/3$,  
is regarded as a first order small quantity.  They have only a small 
effect on the fits of Sec.~IX 
(because for charged fermions the second to third generation mass ratios 
are small), but are useful in performing accurate numerical checks that 
$U_L^f M_f^{\prime}  U_R^{f\dagger}$ is diagonal.

\tightenlines

\begin{table}
\caption[]{Higgs eigenmodes, masses, and fermion couplings for the
3 Higgs doublet model in the cyclic symmetry limit}
\begin{tabular}{c c c c}
mode & charge &  mass & fermion family \\
designation& &  squared & couplings\\  
\hline
$\delta^{(1)}$ & $\pm1$ & $3(\mu_2+\alpha)\Omega^2$ & 1st \\
$\delta^{(2)}$ & $\pm1$ & $3(\mu_2+\alpha)\Omega^2$ & 2nd \\
$\epsilon_R^{(3)}$&0&$4\lambda v^2$& 3rd \\
$\epsilon_R^{(+)}$,$\epsilon_I^{(-)}$&0&$(4\lambda+2\mu_1+2\mu_2+{7\over 
2}\alpha)\Omega^2$&1st and 2nd \\
$\epsilon_R^{(-)}$,$\epsilon_I^{(+)}$&0&${9\over 2}\alpha\Omega^2$&
1st and 2nd \\ 

\end{tabular}
\end{table}

\vspace{30pt}

\begin{table}
\caption[] {Higgs eigenmodes, masses, and fermion couplings for the
6 Higgs doublet model in the cyclic symmetry limit, 
assuming weak coupling of $\phi$ to $\eta$}
\begin{tabular}{c c c c}
mode & charge &  mass & fermion family \\
designation& &  squared & couplings  \\
\hline
$\delta_{\phi}^{(1)}$ & $\pm1$ &
$3(\mu_{2\phi}+\alpha_{\phi})\Omega_{\phi}^2$  & 1st \\
$\delta_{\phi}^{(2)}$ & $\pm1$ &
$3(\mu_{2\phi}+\alpha_{\phi})\Omega_{\phi}^2$  & 2nd \\
$\epsilon^{(3)}_{\phi R}$&0&$4\lambda_{\phi} v_{\phi}^2$& 3rd \\
$\epsilon_{\phi R}^{(+)}$,$\epsilon_{\phi
I}^{(-)}$&0&$(4\lambda_{\phi}+2\mu_{1\phi}+2\mu_{2\phi}+{7\over
2}\alpha_{\phi})\Omega_{\phi}^2$&1st and 2nd \\
$\epsilon_{\phi R}^{(-)}$,$\epsilon_{\phi I}^{(+)}$&0&${9\over
2}\alpha_{\phi}\Omega_{\phi}^2$& 1st and 2nd \\
\hline
$\delta_{\eta}^{(1)}$ & $\pm1$ &
$3(\mu_{2\eta}+\alpha_{\eta})\Omega_{\eta}^2$ & 3rd \\
$\delta_{\eta}^{(2)}$ & $\pm1$ &
$3(\mu_{2\eta}+\alpha_{\eta})\Omega_{\eta}^2$ & 1st \\
$\epsilon^{(3)}_{\eta R}$&0&$4\lambda_{\eta} v_{\eta}^2$& 2nd \\
$\epsilon_{\eta R}^{(+)}$,$\epsilon_{\eta
I}^{(-)}$&0&$(4\lambda_{\eta}+2\mu_{1\eta}+2\mu_{2\eta}+{7\over
2}\alpha_{\eta})\Omega_{\eta}^2$&1st and 3rd \\
$\epsilon_{\eta R}^{(-)}$,$\epsilon_{\eta I}^{(+)}$&0&${9\over
2}\alpha_{\eta}\Omega_{\eta}^2$& 1st and 3rd \\
\hline
$\delta_{PG}^{(3)}$ & $\pm1$&$\sim |V_2|/\Omega^2$& 2nd and 3rd \\
$\epsilon_{PG}^{(3)}$ & 0&  $\sim |V_2|/\Omega^2$& 2nd and 3rd \\

\end{tabular}
\end{table}

\begin{table}
\caption[]{Six Higgs model fit to experimental data}
\begin{tabular}{c c c}
quantity & target value & fitted value\\
\hline
$v=[6(\Omega_{\phi}^2+\Omega_{\eta}^2)]^{1\over 2}$ & 247. & 247. \\
$\Omega_{\phi}-\Omega_{\eta}$ & 0. &0.001 \\
\hline
$M_u$ & 0.005 & 0.005 \\
$M_c$ & 1.30 & 1.28 \\
$M_t$ & 173. & 173. \\
\hline
$M_d$ & 0.010 & 0.011 \\
$M_s$ & 0.200 & 0.219 \\
$M_b$ & 4.30 & 4.29 \\
\hline
$M_e$ & 0.00051 & 0.00051 \\
$M_{\mu}$ & 0.1057 & 0.1057 \\
$M_{\tau}$ & 1.777  & 1.777 \\
\hline
${|D_{\rm Higgs}|\over 2.6\times 10^{-14}}$ & 0. & 0.016\\
\hline
$s_{12}$ & 0.221 & 0.221 \\
$s_{13}$ & 0.0035 & 0.0041\\
$s_{23}$ & 0.041 & 0.035 \\
$|\sin \delta_{13}|$ & 0.60 & 0.44 \\

\end{tabular}
\end{table}

\begin{table}
\caption[]{Three Higgs model fit to experimental data}
\begin{tabular} {c c c}

quantity & target value & fitted value\\
\hline
$v=6^{1\over 2}\Omega_{\phi}$ & 247. & 247. \\
\hline
$M_u$ & 0.005 & 0.005 \\
$M_c$ & 1.30 & 1.28 \\
$M_t$ & 173. & 173. \\
\hline
$M_d$ & 0.010 & 0.011 \\
$M_s$ & 0.200 & 0.037 \\
$M_b$ & 4.30 & 4.31 \\
\hline
$M_e$ & 0.00051 & 0.00051\\
$M_{\mu}$ & 0.1057 & 0.1057 \\
$M_{\tau}$ & 1.777  & 1.777 \\
\hline
${|D_{\rm Higgs}|\over 2.6\times 10^{-14}}$ & 0. & 0.001  \\
\hline
$s_{12}$ & 0.221 & 0.221 \\
$s_{13}$ & 0.0035 & 0.0037 \\
$s_{23}$ & 0.041 & 0.039 \\
$|\sin \delta_{13}|$ & 0.60 & 0.55 \\

\end{tabular}
\end{table}

\end{document}